\documentclass[english,12pt]{article}
\usepackage[T1]{fontenc}
\usepackage[latin9]{inputenc}
\usepackage{geometry}
\usepackage{color}
\usepackage{calc}
\usepackage{amsmath}
\usepackage{cancel}
\usepackage{graphicx}
\usepackage{esint}
\usepackage{comment}
\usepackage{amsfonts}
\usepackage{amsthm}
\usepackage[titletoc,title]{appendix}
\usepackage{mathtools}
\usepackage{afterpage}

\usepackage{setspace} 

\usepackage{amssymb}
\usepackage{bbold}

\makeatletter

\usepackage{slashed}
\usepackage{physics}
\usepackage{tikz-cd}

\usepackage{hyperref}
\hypersetup{
	colorlinks,
	citecolor=blue,
	filecolor=blue,
	linkcolor=blue,
	urlcolor=blue,
	linktocpage=true
}

\makeatother

\usepackage[backend=bibtex,sorting=none,style=numeric-comp,url=false]{biblatex}
\renewbibmacro{in:}{}
\usepackage{babel}
\bibliography{refs}{}

\numberwithin{equation}{section}

\newcommand{\twomatrix}[4]{\begin{pmatrix} {#1} & {#2} \\ {#3} & {#4} \end{pmatrix}}

\newcommand{\C}{\mathcal{C}}
\newcommand{\Z}{\mathbb{Z}}
\newcommand{\R}{\mathbb{R}}
\renewcommand{\P}{\mathcal{P}}
\newcommand{\omegab}{{\overline{\omega}}}

\renewcommand{\mod}{{\;(\text{mod})}}

\usepackage{amsthm}
\newtheorem{innercustomgeneric}{\customgenericname}
\providecommand{\customgenericname}{}
\newcommand{\newcustomtheorem}[2]{%
	\newenvironment{#1}[1]
	{%
		\renewcommand\customgenericname{#2}%
		\renewcommand\theinnercustomgeneric{##1}%
		\innercustomgeneric
	}
	{\endinnercustomgeneric}
}

\newcustomtheorem{customthm}{Theorem}
\newcustomtheorem{customclaim}{}
\newcustomtheorem{customcorollary}{Corollary}
\newcustomtheorem{customconstruction}{Construction}

\usepackage{authblk}

\begin{document}

\title{Non-rational Narain CFTs from codes over $F_4$}
\author[1,2]{Anatoly Dymarsky}
\author[3]{Adar Sharon}

\affil[1]{\small{Department of Physics and Astronomy, University of Kentucky, Lexington, KY 40506}}
\affil[2]{\small{Skolkovo Institute of Science and Technology, Skolkovo Innovation Center, Moscow, Russia, 143026}}
\affil[3]{\small{Department of Particle Physics and Astrophysics,
	Weizmann Institute of Science, Rehovot, Israel, 7610001}}

\maketitle

\abstract{  
We construct a map between a class of codes over $F_4$ and a family of non-rational Narain CFTs. This construction is complementary to a recently introduced relation between quantum stabilizer codes and a  class of rational Narain theories. From the modular bootstrap point of view we formulate a polynomial ansatz for the partition function which reduces modular invariance to a handful of algebraic easy-to-solve constraints. For certain small values of central charge our construction yields optimal theories, i.e.~those with the largest value of the spectral gap.}

\setcounter{tocdepth}{2}
\tableofcontents


\section{Introduction}\label{sec:intro}
Narain theories, two-dimensional conformal theories of free fields compactified on a multi-dimensional torus, have enjoyed renewed attention recently. Both in the context of holographic correspondence and the modular bootstrap program, Narain theories play the  Goldilocks role of theories rich enough and simple enough to be studied.  Holographically, the emphasis is on the bulk description of the ensemble of Narain theories \cite{Afkhami-Jeddi:2020ezh,maloney2020averaging,perez2020gravitational,datta2021adding,benjamin2021narain,Ashwinkumar:2021kav,Dong:2021wot,Collier:2021rsn}. As a parallel development, the spectral gap of Narain theories has being studied numerically using the modular bootstrap approach \cite{hartman2019sphere,afkhami2020high,Afkhami-Jeddi:2020ezh}. In both cases, the hope is that lessons learned for Narain theories, which exhibit $U(1)^n \times U(1)^n$ symmetry, will be relevant for the conventional ``Virasoro'' CFTs without extended symmetries. 

Two years ago spinless bootstrap constraints for theories with $U(1)^n \times U(1)^n$ symmetry were shown in \cite{hartman2019sphere} to reduce  to  linear programming bounds of Cohen and Elkies  on the density of sphere packings \cite{cohn2003new}. This remarkable result establishes a connection between 2d CFTs and a well-known problem in discrete mathematics. More recently, a certain family of Narain theories was found to be related to quantum stabilizer codes \cite{dymarsky2021solutions,Dymarsky:2020qom}. These developments are conceptually similar. First, in both cases a subset of modular bootstrap constraints reduces to a well-known problem, the linear programming bounds of \cite{cohn2003new} in the case of sphere packings and  those of Calderbank et al.~in the case of quantum codes \cite{calderbank1998quantum}. Second, the problems of maximizing the CFT spectral gap, sphere packing density, and code Hamming (or other appropriate) distance are qualitatively similar, which can be utilized e.g.~to shed light on holographic properties of Narain theories \cite{calderbank1998quantum}. We further elucidate this point below.

In the context of classical codes, a central unsolved problem is of finding codes $[n,k,d]$ of fixed length $n$ and rank (encoding code capacity) $k$  with the maximal possible Hamming distance $d$. Such codes are called {\it optimal}. The asymptotic value of maximal $d/n$ for fixed $k/n$ and $n\rightarrow \infty$ are not known. Maximal $d$ is constrained by linear programming bounds. If a code saturates a linear programming bound it is called {\it extremal} \cite{nebe2006self}. An extremal code is automatically optimal but not vise versa. Historically in the context of double-even self-dual binary codes, for which $k=n/2$, extremal codes were defined as those saturating a particular analytic linear programming bound $d= 4[n/24]+4$. As the linear programming bounds improved, some authors call the code extremal if it saturates any subset of necessary constraints -- the nomenclature we follow. It is an open question if the optimal code(s) for $n\gg 1$, say optimal double-even self-dual linear binary code(s) with $n=72$, are extremal. A very similar situation and nomenclature applies to quantum codes. 

\afterpage{\begin{table}
		\centering
		$$
		\begin{array}{l|cc|cc}
		c & \Delta_{1}  & \text { theory/lattice} & \text {$\Delta_1$} & \text {lattice} \\
		\hline 1 & 1 / 2  & S U(2)_{1} \text { WZW } & \sqrt{1 / 3} & A_2 \\
		2 & 2 / 3 & S U(3)_{1} \text { WZW } & \sqrt{1 / 2} & D_4  \\
		3 & 3 / 4  & S U(4)_{1} \text { WZW } & \sqrt[6]{1 / 3} & E_6  \\
		4 & 1  & S O(8)_{1} \text { WZW } & 1  & E_8\\
		5 & 1  & S O(10)_{1} \text { WZW } & \sqrt[10]{4 / 3} & \Lambda_{10} \\
		6 & \sqrt{4 / 3}  & \text { Coxeter-Todd } & \sqrt{4 / 3} & \text { Coxeter-Todd }  \\
		7 & \sqrt{4 / 3}  & & \sqrt[14]{64 / 3} & \Lambda_{14} \\
		8 & \sqrt{2} & \text { Barnes-Wall } & \sqrt{2} & \text { Barnes-Wall }  \\
		\end{array}
		$$
		\caption{Table of hypothetical optimal Narain theories \cite{Afkhami-Jeddi:2020ezh} (left panel) and densest lattice sphere pickings in ${\mathbb R}^{2c}$ \cite{SPLAG} (right panel). Density of sphere packings in measured in $\Delta_1$, which is half of the shortest vector length squared. }
		\label{listofoptimal}
\end{table}}
For Narain CFTs, we propose to call a theory optimal if it maximizes the spectral gap for the given central charge $c$ and extremal if it saturates (any subset) of the modular bootstrap constraints. A conjectural list of optimal theories for  $1\leq c\leq 8$ was given in \cite{Afkhami-Jeddi:2020ezh}, together with optimal (densest) {\it lattice} sphere pickings in ${2c}$ dimensions, see table \ref{listofoptimal}.
A brief examination reveals that the picture for maximal spectral gap and sphere pickings is similar -- for certain small dimensions optimal lattices are related to codes. This is most known for the densest sphere pickings in $8$ and $24$ dimensions, which are related to $E_8$ and Leech lattices (Hamming $[8,4,4]$ and Golay $[24,12,8]$ codes) correspondingly. But in fact for other dimensions optimal lattices are related to codes as well\footnote{For example, lattice $D_4$ which is optimal in ${\R}^4$, is the construction A lattice of the linear binary code consisting of all codewords of even weight; the Coxeter-Todd lattice optimal in ${\R}^6$ is the construction A lattice of the Hexacode, as is reviewed in section \ref{sec:hexacode}.} \cite{SPLAG}. For the Narain case, conjectural optimal theories with $c=3,4,5$ from the Table \ref{listofoptimal} are the code CFTs associated with quantum stabilizer codes of \cite{dymarsky2021solutions,Dymarsky:2020qom}. Up to T-dualities such codes/CFTs can be parametrized by graphs on $c=n$ nodes. For $c=3,4,5$ the optimal theories are code CFTs associated with the fully connected graphs. Among them is the $c=4$ theory, associated with the $E_8$ lattice understood as a Narain lattice, which saturates the numerical Virasoro bootstrap constraints and hence it is extremal and optimal among {\it all} 2d CFTs \cite{Collier:2016cls,hartman2019sphere,Afkhami-Jeddi:2020ezh}.

All optimal theories from  table \ref{listofoptimal} exhibit  ``quantized'' spectrum, i.e.~the conformal dimensions of $U(1)^c\times U(1)^c$ primaries are integer in some appropriate units,
\begin{equation*}
    \Delta=k\, \Delta^*, \qquad k \in \Z^+,
\end{equation*}
where $\Delta^*$ may be irrational. This hints optimal theories beyond $c=3,4,5$ might be related to codes as well. In this paper we introduce a novel construction, complementary to the construction of \cite{dymarsky2021solutions,Dymarsky:2020qom} which maps a certain class of isodual codes over $F_4$, which we call codes of N-type, to a family of Narain CFTs. We will call these theories code CFTs. The spectrum gap of code CFTs is quantized in the units of $\Delta^*=1/\sqrt{3}$ which implies that these are non-rational theories \cite{vafa1988toward}. This is quite remarkable given that code CFT partition functions exhibit a simple algebraic structure, hinting at a finite number of  ``characters.'' 
The spectral gap of a code CFT associated with the code $\cal C$ is given by
\begin{equation}
    \Delta_1=\frac{\min\{d,4\}}{2\sqrt{3}}, \label{sg}
\end{equation}
where $d$ is the Hamming distance of $\C$. Using this construction, we will show optimal Narain theories from table \ref{listofoptimal} with $c=6,7$ are code CFTs. 

Similarly to the case of \cite{dymarsky2021solutions,Dymarsky:2020qom} where the CFT torus partition function had a closed-form expression in terms of the code's enumerator polynomial, in the present case the partition function can be expressed in terms of {\it extended} enumerator polynomial $W^S$, defined in equation \eqref{eq:extended_enumerator},
\begin{equation}
    Z={W^S(\{\psi_x\},\{\psi_{xy}\})\over |\eta|^{2n}}.
\end{equation}
Here $\psi_x$ and $\psi_{xy}$ are particular Seigel theta functions defined in \eqref{eq:psi_i} and \eqref{eq:psi_ij}.  
Modular invariance of $Z$ is guaranteed by the algebraic properties of $W^S$, outlined in \eqref{eq:mac}, \eqref{eq:T_transf}. In fact any polynomial $W_S$ satisfying those properties gives rise to a modular invariant $Z$, serving as a useful ansatz solving the modular bootstrap constraints.

The paper is organized as follows. In section \ref{sec:background} we remind the reader the basics of  codes and their relation to lattices. In section \ref{sec:construction} we define the codes of N-type and construct an explicit map from these codes to Narain lattices. In section \ref{sec:analyzing_code_CFTs} we parametrize all N-type codes and calculate their Seigel theta series in terms of the extended enumerator polynomial $W^S$. We also discuss algebraic symmetries of $W^S$ and how they are solved  by a ring of invariant polynomials. Section \ref{sec:simple_examples} lists explicit examples, and in section \ref{sec:constructing_optimal} we construct the optimal theories for $c=6,7$ using N-type codes. We conclude in section \ref{sec:conclusions}.

\section{Background}
\label{sec:background}

In this section we review relevant background for codes over $F_4$ and Narain CFTs. For a more detailed pedagogical introduction see \cite{Dymarsky:2020qom}.

\subsection{Codes and lattices}

A linear binary code $\C$ is a $k$-dimensional vector space in $F_2^n$ over the field $F_2$. The field $F_2=\{0,1\}$ consists of two elements with the conventional operations and $1+1=0$.
In other words $\C$ is a 
set of $2^k$ binary vectors, also called codewords.  Elements of codewords $c\in \C$ are called letters. 
The vector space $F_2^n$ is equipped with the Hamming norm (weight) $w(c)$ which evaluates the number of non-zero letters of $c \in \C \subset F_2^n$. The Hamming distance of a code is defined as the minimal Hamming weight of all of its non-trivial codewords, 
\begin{equation}
d=\min_{c\in\C,\;c\neq 0^n} w(c)\;.
\label{Hd}
\end{equation}
A code $\C\subset F_2^n$ of size $2^k$ and Hamming distance $d$ is said to be of type $[n,k,d]$.

We define an inner product $(\cdot,\cdot)_B$ for binary codewords in the obvious way:
\begin{equation}\label{eq:binary_product}
(c_1, c_2)_B=\sum_{i=1}^n c^i_1  c^i_2 \mod\, 2,\qquad c_1,c_2\in F_2^n\;.
\end{equation}
This allows us define a dual code $\C^*$ as the vector subspace in $F_2^n$ orthogonal to $\C$, i.e.~consisting of the binary codewords which are orthogonal to every codeword of $\C$.
A code is self-orthogonal if $\C^*\subset \C$, and self-dual if $\C^*= \C$. 
The simplest example of a self-dual code is the $[2,1,2]$ ``repetition'' code consisting of two codewords, the trivial one $c_0=(0,0)$ and $c_1=(1,1)$.

A mapping from a code to a lattice is called a Construction. The most basic one is called Construction A: for additive codes over $F_2$ the lattice $\Lambda(\C)$ associated with the code $\C$ 
is  given by
\begin{equation}\label{eq:construction_A_binary}
\Lambda(\C)=\{\vec v/\sqrt 2\, |\, \vec v\in(F_2)^n,\;\vec v\equiv c\mod\, 2,\;x\in \C \}\subset \mathbb{R}^n\;.
\end{equation}
The normalization by $\sqrt 2$ is necessary to ensure  the lattice of the dual code is just the dual lattice,
\begin{equation}\label{eq:fund_relation_binary}
\Lambda(\C^*)=(\Lambda(\C))^*\;.
\end{equation} 
Certain properties of the code $\C$ translate to the properties of the lattice $\Lambda(\C)$. In particular, if $\C$ is a self-dual code, then  $\Lambda(\C)$ is a self-dual lattice. In addition, if $\C$ is an even code, then every vector of $\Lambda(\C)$ has integral norm-squared.

To proceed  further we define $F_4$, the field of four elements $\{0,1,\omega, \omegab\}$. It satisfies the conventional algebra $0\, x=0$, $x+0=0$, and $1\, x=x$ for any $x\in F_4$. In addition, it also satisfies $x+x=0$; the sum of any two non-zero elements is equal to third one: $1+\omega=\bar \omega$, $1+\bar \omega=\omega$, $\omega+\bar\omega=1$; and $\omega\, \bar \omega=1$. $F_4$ admits an external automorphism conjugation which exchanges $\omega \leftrightarrow \omegab$.

An additive code over $F_4$ is defined as a $k$-dimensional vector subspace $\C\in F_4^n$. It consists of $2^k$ codewords and is denoted $[n,k,d]$, where the Hamming distance $d$ is defined as in \eqref{Hd} with the Hamming weight $w(c)$ counting the total number of non-zero elements of $c\in \C$. There are many different ways to define an inner product on $F_4^n$. Throught the paper we will use the Hermitian inner product
\begin{equation}\label{eq:hermitian_product}
(c_1, c_2)=\sum_{i=1}^n \overline{c}^i_1 c^i_2 +
c_1^i\overline{c}^i_2,\qquad c,c'\in\C\;.
\end{equation}
All algebra is understood in the sense of $F_4$ and therefore $(c_1, c_2)$ is either zero or one. 
As in the case of binary codes, for codes over $F_4$ a dual code $\C^*$ is defined as the vector subspace in $F_4^n$ orthogonal to $\C$ with respect to the inner product \eqref{eq:hermitian_product}. For a $[n,k,d]$ code $\C$, the dual code $\C^*$ would be $[n,2n-k,\tilde{d}]$.
The simplest example of a self-dual code would be the $[1,1,1]$ code consisting of a trial codeword $C_0=(0)$ and $c_1=(x)$
where $x$ is either $\omega,\bar\omega$ or $1$. 

The algebraic properties of $F_4$ become apparent if we identify elements $x\in F_4$ with points on the complex plane. Namely $0,1\in F_4$ are mapped to $0,1\in \mathbb C$, while $\omega, \bar\omega$ are mapped to $e^{\pm 2\pi i/3}$. Upon imposing the equivalence condition $2\,x=0$ we obtain $F_4$.
This simple observation is the idea behind Construction A, which is a mapping from codes over $F_4$ to lattices. First, any element $x\in F_4$ can be represented as $x=a\, \omega+b\,\omegab$ where $a,b\in \Z_2$. This defines an invertible map, called Gray map, from $F_4 \rightarrow F_2^2$ and codewords $c\in \C$ can be represented as binary vectors in $F_2^{2n}$. The lattice associated with $\C$ is defined as
\begin{equation}
\Lambda(\C)=\{\vec a\,\omega+\vec b\, \bar \omega\, |\;(\vec a,\vec b) \mod\, 2 \equiv c \in \C \}\subset \mathbb{C}^n=\mathbb{R}^{2n}\;, \quad \omega=e^{2\pi i/3},\quad \bar \omega=e^{-2\pi i/3}.
\label{LC}
\end{equation}

For a self-dual code $\C$, the corresponding lattice \eqref{LC} is 3-modular \cite{nebe2006self}. If we further rescale it by $1/3^{1/4}$, the resulting lattice 
\begin{equation}
\tilde \Lambda(\C)=\Lambda(\C)/3^{1/4}
\label{LC3}
\end{equation}
will be iso-dual, in the sense that the dual lattice is related to $\tilde \Lambda(\C)$ by a $\pi/2$ rotation in each complex plane of $\mathbb{C}^n$ \cite{Dymarsky:2020qom}. 
More generally, for an arbitrary code (see appendix \ref{app:isoduality} for a proof), 
\begin{equation}\label{eq:isodual}
\tilde \Lambda^*(\C)=O_{\pi/2}^{(n)}\tilde \Lambda(\C^*)\, .
\end{equation}
Here $O_{\pi/2}^{(n)}$  acts in each complex plane $\mathbb C=\mathbb R^2$ with 
\begin{equation}
O_{\pi/2}= \begin{pmatrix}0 & 1 \\ -1 & 0 \end{pmatrix}
\end{equation} 
and in \eqref{eq:isodual} we abuse the notation by denoting both the lattice and its generating matrix by 
$\Lambda(\C)$. It should be noted that lattice generating matrix is not unique, an action by 
${\rm GL}(2n,\Z)$ defines the same lattice. Hence \eqref{eq:isodual} could be understood either in the sense of the equivalence under ${\rm GL}(2n,\Z)$ or that one can find representatives $\Lambda^*(\C)$ 
and $\Lambda(\C^*)$ satisfying \eqref{eq:isodual} as the matrix identity.

In the rest of the paper we will use somewhat different coordinates to represent  $\tilde{\Lambda}(\C)\subset \R^{2n}$. The first $n$ coordinates will be the $x$-coordinates of the $n$ complex planes of \eqref{LC}, while the last $n$ coordinates will be the $y$-coordinates, 
\begin{equation}
\label{CA}
\tilde{\Lambda}(\C)=\left.\left\{{(\Re(\vec a\,\omega+\vec b\, \bar \omega),\Im(\vec a\,\omega+\vec b\, \bar \omega))\over 3^{1/4}}\, \right| \;(\vec a,\vec b) \mod\, 2 \equiv c \in \C \right\}\subset \mathbb{R}^{2n}\; .
\end{equation}

\subsection{Narain CFTs}

Narain CFTs are theories describing compactificsation of $n$ free scalars on a torus parametrized by a metric $G$ and a $B$-field. Mathematically, each Narain theory is uniquely specified by a Narain lattice $\Lambda$, which is an even self-dual lattice in $\mathbb R^{n,n}$. With the conventional metric 
\begin{equation}
\eta = \twomatrix{\mathbb{1}_n}{}{}{-\mathbb{1}_n}\;
\end{equation}
lattice vectors are usually denoted as $(p_L,p_R)\in \Lambda$.
The partition function of a Narain CFT on a Euclidean torus $\tau$ is given by
\begin{equation}\label{eq:narain_partition}
Z(\tau,\bar{ \tau})=\frac{1}{|\eta(\tau)|^{2n}}\sum_{(p_L,p_R)\in\Lambda} q^{p_L^2/2}\overline{q}^{p_R^2/2}\;,\quad q=e^{2\pi i\tau},\quad \overline q=e^{-2\pi i\bar \tau}\;.
\end{equation}
Clearly orthogonal transformations $O(n)\times O(n)\subset O(n,n)$ which individually rotate $p_L$ and $p_R$ do not change the theory, although they change the lattice. 

It is convenient to introduce coordinates to $(\alpha,\beta)$ defined by
\begin{equation}
\alpha=\frac{p_L+p_R}{\sqrt 2},\quad \beta=\frac{p_L-p_R}{\sqrt 2}\;,
\end{equation}
In these coordinates the metric becomes
\begin{equation}\label{eq:metric_g}
g=\twomatrix{}{\mathbb{1}_n}{\mathbb{1}_n}{}\;.
\end{equation}
It is always possible to apply $O(n)\times O(n)$ transformations and choose a
lattice generator matrix in the form (here and below we use $\Lambda$ to denote both the lattice and its generator matrix)
\begin{equation}\label{eq:Lambda}
\Lambda=\twomatrix{\gamma^*}{B\gamma}{}{\gamma}\;,
\end{equation}
where $\gamma^*=(\gamma^{-1})^{T}$ is the dual lattice to $\gamma$. The matrix $\gamma$ generates the lattice which defines the compactification torus, $G=\gamma^T\gamma$ while $B$ is an antisymmetric $B$-field. 
The form of $\Lambda$ \eqref{eq:Lambda} is convenient because it manifests self-duality of $\Lambda$; as a matrix $\Lambda\in O(n,n)$ and obeys 
\begin{equation}
\Lambda^Tg\Lambda=g \in {\rm GL}(2n,\Z)\;,
\end{equation}
and so the lattice is self-dual with respect to the metric $g$.

The central question of the modular bootstrap program is to determine the maximal value of the spectral gap among all theories with a particular value of the central charge $c$. For Narain theories, the question is about $U(1)^n \times U(1)^n$ primaries, i.e.~the length squared of the shortest non-trivial lattice vector \begin{equation}
    \Delta_1=\min_{\substack{(p_L, p_R)\in \Lambda,\\ p_L^2+p_R^2\neq 0}} {p_L^2+p_R^2\over 2}\;,
\end{equation}
which one would like to maximize among all Narain lattices of the same dimension $2n=2c$.

\section{Constructing Narain CFTs from codes over $F_4$}\label{sec:construction}

In this section we describe construction \ref{constructionN}, 
which maps certain codes over $F_4$ to Narain CFTs. As discussed in the section above, any code $\C$ over $F_4$ can be mapped to a lattice $\tilde{\Lambda}(\C)$ via \eqref{LC3}. We might then expect that provided certain conditions on the code $\C$ are satisfied, the resulting lattice will be a Narain lattice with respect to a properly defined Lorentizan metric.

\subsection{General construction}

We start by discussing the most general case. 
Assume $\C$ is an additive code over $F_4$ of length $n$, and define its conjugate $\overline {\C}$ to be the code $\C$ with each letter $\omega$ replaced by $\omegab$ and vice-versa. Also define $\tilde{\Lambda}(\C)\subset\mathbb{R}^{2n}$ to be the lattice obtained via construction A from $\C$ and rescaled by $1/3^{1/4}$ \eqref{CA}.

An important element of the construction will be a permutation of the letters of the code. Define $S$ to be a permutation of the letters which only permutes letters in pairs, $S^{-1}=S$.  We can think of $S$ as an $n\times n$ orthogonal integral matrix, which then obeys $S=S^T=S^{-1}$. We denote by $S(\C)$ the code $\C$ where we have permuted the letters of each codeword according to $S$.
We now claim the following.
\begin{customclaim}{Self-duality}\label{claim1}
	If $\C^*=S(\overline \C)$, then $\tilde \Lambda(\C)$ is self-dual with respect to the metric
	\begin{equation}\label{eq:metric_gS}
	 g_S=\twomatrix{}{1}{1}{}\otimes S =\twomatrix{}{S}{S}{}\;.
	\end{equation} 
	\end{customclaim}
\begin{customclaim}{Evenness}\label{claim2}
	  Each letter $c^i$ of a codeword $c\in\C$ can be written as $c^i=a^i\,\omega + b^i\, \overline \omega$ for $a^i,b^i\in\Z_2$. Then if 
	\begin{equation}\label{eq:eveness_condition}
	\sum_{i=1}^n \left(b^i b^{S(i)}-a^i a^{S(i)}\right) \mod\, 4 =0
	\end{equation}
	for every $c\in\C$, then the lattice $\tilde \Lambda(\C)$ is even with respect to the metric $g_S$. 
\end{customclaim}
To use these claims to construct a Narain CFT from a code over $F_4$ it would be enough to show that $g_s$ is a Lorentizan metric with the signature $(n,n)$ in $\mathbb R^{2n}$. Indeed, $g_S$ is related to \eqref{eq:metric_g} by a similarity transformation, $g_S=O^TgO$, with the orthogonal matrix $O=\twomatrix{\mathbb{1}_{n}}{}{}{S}$. The lattice $O \tilde{\Lambda}(\C)$ is then even and self-dual with respect to $g$, and hence defines a Narain CFT.
$\;$\\
\begin{customconstruction}{N}\label{constructionN}
	If $\C$ obeys $\C^*=S(\overline \C)$ and each codeword $c\in\C$ obeys \eqref{eq:eveness_condition}, then $\tilde{\Lambda}(\C)=\Lambda(\C)/3^{1/4}$ is a Narain lattice with respect to $g_S$.
\end{customconstruction}
$\;$\\
Construction \ref{constructionN} specifies a certain class of codes over $F_4$ satisfying  \ref{claim1} and \ref{claim2}. We will call such codes  ``N-type codes'', and corresponding CFTs obtained via this construction ``code CFTs''. Since permutation and conjugations are transformations from the code automorphism group \cite{calderbank1998quantum}, N-type codes are isodual $[n,n,d]$ codes over $F_4$.   

Before we proceed to prove the claims, let us explain why the pairwise permutation $S$ can not be easily generalized to an arbitrary permutation. The condition $S^2=1$ is crucial for the metric \eqref{eq:metric_gS} to satisfy $g_S^2=1$, a necessary condition provided $g_S$ is related to $g$ by an orthogonal transformation. 

Let us prove the claims. We start with \ref{claim1}. Following \eqref{eq:isodual}, the dual lattice $\tilde\Lambda^*$ is related to $\tilde\Lambda$ by
\begin{equation}\label{eq:2}
	\tilde\Lambda^*(\C)=O_{\pi/2}^{(n)} \tilde\Lambda(\C^*)\;.
\end{equation}
We will abuse notation and denote by $\tilde\Lambda$ also the generator matrix for the lattice $\tilde\Lambda$, in which case $\tilde\Lambda^*=(\tilde\Lambda^{-1})^{T}$. Then we can interpret \eqref{eq:2} as matrix multiplication.

Since $\C^*=S(\overline\C)$, 
we find that 
\begin{equation}
\tilde\Lambda^*(\C)=O_{\pi/2}^{(n)}P^{(n)}\sigma  \tilde \Lambda(\C)\;,
\end{equation}
where 
\begin{equation}
P^{(n)}=\twomatrix{-1}{}{}{1}\otimes\mathbb{1}_n=\twomatrix{-\mathbb{1}_n}{}{}{\mathbb{1}_n}\;,\qquad \sigma=\mathbb{1}_2\otimes S=\twomatrix{S}{}{}{S}\;.
\end{equation}
Specifically, $P^{(n)}$ performs a conjugation in each complex plane, while $\sigma$ interchanges $n$ complex planes according to the permutation $S$. Note that
\begin{equation}
O_{\pi/2}^nP^n=g=\twomatrix{}{\mathbb{1}_{n}}{\mathbb{1}_{n}}{}\;,
\end{equation} 
and that $g$ commutes with $S$. Then we have found 
\begin{equation}\label{eq:eq1}
\tilde\Lambda^*(\C)=g_S\,  \tilde \Lambda(\C)\;,
\end{equation}
where $g_S=\sigma g=\twomatrix{}{S}{S}{}$. Equation \eqref{eq:eq1} manifests that $\tilde\Lambda(\C)$ is self-dual with respect to $g_S$.

Next we prove \ref{claim2}. The norm of a lattice vector $\vec v\in\tilde\Lambda(\C)$ is 
\begin{equation}
|\vec v|^2= v^T g_S  v\;.
\end{equation}
Vector $v$ can be also written as an $n$-dimensional complex vector with the coordinates  $v^i=\frac{a^i\,\omega+b^i\,\omegab }{3^{1/4}}$ for $a^i,b^i\in\mathbb{Z}$ such that 
\begin{equation}\label{eq:norm}
|\vec v|^2=\frac{1}{2}  \sum_{i=1}^n \left(b^i b^{S(i)}-a^i a^{S(i)}\right)\;.
\end{equation}
It is enough to show that $|\vec v|^2 \mod\, 2=0$. First we show that we can restrict to $a_i,b_i\in\{0,1\}$. Indeed, shifting $a^i\to a^i+2$ for a particular $i$ we find 
\begin{equation}
|\vec v|^2\to |\vec v|^2+a^{S(i)}+a^{S^{-1}(i)}\equiv |\vec v|^2\mod\, 2 \;,
\end{equation}
where we used $S=S^{-1}$. A similar argument works for $b^i$. However, if $a^i,b^i\in\{0,1\}$, then $ v$ is also a codeword of $\C$. It is thus enough to have 
\begin{equation}
\sum_{i=1}^n \left(b^i b^{S(i)}-a^i a^{S(i)}\right)\equiv 0 \mod\, 4
\end{equation}
for any codeword $c$, which is exactly our assumption \eqref{eq:eveness_condition}.

\subsection{Simplified construction for self-dual codes}\label{sec:simple_construction}
The general construction simplifies for self-dual codes $\C=\C^*$. Then the condition described above  reduces to the following one:
$\;$\\
\begin{customconstruction}{N$'$}\label{constructionNp}
	Consider a self-dual code $\C$. For every $c\in \C$, define the weight $w_{x}(c)$ for $x\in F_4$ to be the number of letters in $c$ which are equal to $x$. Then if for all codewords $c\in \C$ satisy
	\begin{equation}\label{eq:self_dual_assumption}
	w_{\overline\omega}(c)-w_{\omega}(c)\equiv 0\mod\, 4\;,
	\end{equation} 
	the lattice  $\tilde{\Lambda}(\C)=\Lambda(\C)/3^{1/4}$ is a Narain lattice with respect to the Lorentzian metric $g$.
\end{customconstruction} 
$\;$\\
The construction \ref{constructionNp} is equivalent to construction \ref{constructionN} (with a trivial permutation $S=1$) when it applies. To prove the claim we first show that $\tilde{\Lambda}$ is even. The condition \ref{claim2} in this case reduces to 
\begin{equation}
\sum_{i=1}^n \left((b^i)^2 -(a^i)^2\right)\equiv 0 \mod\, 4\;,
\label{a2b2}
\end{equation}
for all codewords. Since $a^i,b^i\in\{0,1\}$, we can set $(a^i)^2=a^i$ and similarly for $b^i$. Also note, if the letter $c^i$ is equal to $1$, it has $a_i=b_i$ and so it does not contribute to this sum, so that \eqref{a2b2} can equivalently be reformulated as \eqref{eq:self_dual_assumption}. 
To finish the proof, we must show  the lattice is also self-dual with respect to $g$. 
Since $\C$ is self-dual, to 
reduce it to  \ref{claim1}, it would  be enough to show that $\C=\overline{\C}$. Indeed it can be shown that any self-dual code that obeys \eqref{eq:self_dual_assumption} is invariant under conjugation, see appendix \ref{app:proof}.

\section{Analyzing code CFTs}\label{sec:analyzing_code_CFTs}

\subsection{Universal properties of code CFTs}

There are universal properties which are common to all N-type codes and code CFTs obtained via construction \ref{constructionN}. We describe them below.

\subsubsection{The binary subcode}\label{sec:binary_subcode}

For any code $\C$ over $F_4$, we can define its binary subcode $\C_B\subset \C$ which consists of only those codewords whose letters are either $0$ or $1$, so that
\begin{equation}
\C_B=\{c\,|\, c\in \C,\,  c\in F_2^n\}\;.
\end{equation}
For N-type codes, this will be some linear $[n,k,d_B]$ binary code, for some $k,d_B\in\mathbb N$. This subcode is usually much easier to study than the full code, and so it is useful to discuss its properties in some detail.

As a reminder, we use $(\cdot,\cdot)$ to denote the standard Hermitian inner product over $F_4$ in equation \eqref{eq:hermitian_product} and $(\cdot,\cdot)_B$ to denote the standard binary inner product over $F_2$ in equation \eqref{eq:binary_product}. In addition, we will define an additional inner product $(\cdot,\cdot)_{B,S}$ which denotes the binary inner product combined with the permutation $S$, i.e.~  $(c_1,c_2)_{B,S}=(c_1,S(c_2))_{B}$ for binary codewords $c_1,c_2\subset F_2^n$.

Now consider some N-type code $\C\subset F_4^n$, with $\C_B$ being its binary subcode. Denote by $\C_B^*$ its binary dual with respect to the standard binary inner product $(\cdot,\cdot)_B$. Then
\begin{equation}
\C_B^*=\{ S(c+\overline c)\, |\, c\in \C \}=\{ c'+\overline c'\, |\, c'\in \C^* \}\;.
\end{equation}
we prove this equation in appendix \ref{app:binary_1}.
It can  be also shown  $\C_B^*$ is even with respect to the inner product $(\cdot,\cdot)_{B,S}$, see appendix \ref{app:binary_2}.

We can use this result to show that a certain codeword must always be part of $\C$. By rearranging letters in our code we can always bring the pairwise permutation $S$ to  the form
\begin{equation}\label{eq:perm_matrix}
S=\overset{m}{\left(\overbrace{\begin{array}{ccc}
		1\\
		& \ddots\\
		&  & 1\\
		\\
		\\
		\\
		\\
		\end{array}}\right.}\overset{n-m}{\left.\overbrace{\begin{array}{ccc}
		\\
		\\
		\\
		\\
		\sigma_{x}\\
		& \ddots\\
		&  & \sigma_{x}
		\end{array}}\right)}\;,
\end{equation}  
where $\sigma_x=\twomatrix{}{1}{1}{}$. $S$ acts trivially on the first $m$ letters and permutes the last $n-m$ letters in pairs ($n-m$ must be even). Now, consider the codeword $\beta=(1^m,0^{n-m})$. Note that since $\C_B^*$ is even with respect to the inner product $(\cdot,\cdot)_{B,S}$, this implies $(\beta,c)_B\equiv 0\mod\, 2$ for every $c\in\C_B^*$. Therefore $\beta\in \C_B\subset \C$, and so $\beta=(1^m,0^{n-m})$ must always be a codeword of our code $\C$.

In the simpler case when we use construction \ref{constructionNp}, the code $\C$ always contains the codeword $b=(1^n)$, so that $\C_B^*$ is even. In addition, in this case the dual binary subcode is self-orthogonal with respect to the usual inner product $(\cdot,\cdot)_B$, so that $\C_B^*\subset \C_B\subset \C$.

\subsubsection{Bounds on the spectral gap}\label{sec:bound_spectral_gap}

There are universal bounds on the spectral gap of any lattice obtained via construction N from a code $\C\subset F_4^n$. Since the Narain lattice is given by a construction A lattice from some code (up to a rescaling by $3^{1/4}$), the spectral gap is always given by \eqref{sg}:
\begin{equation}
\Delta_1=\frac{\min\{d,4\}}{2\sqrt 3}\;.
\end{equation}
For example, the lattice always includes the vector $(2,0,...,0)/3^{1/4}$, and so $\Delta_1\leq \frac{4}{2\sqrt 3}$. 

For codes with large $n$, it is hard to find the Hamming distance $d$. However, there are simpler bounds on $\Delta_1$ which can be obtained. For example, the binary subcode of $\C$ is usually much simpler than $\C$ itself, and $d_B=d(\C_B)$ is an upper bound on the hamming distance
\begin{equation}
    \Delta_1\leq \frac{d(\C_B)}{2\sqrt 3}.
\end{equation} 
Similarly, we found above that if the permutation $S$ keeps $m>0$ letters invariant, the code includes a codeword with $m$ ones and zeros otherwise. This gives an upper bound on $\Delta_1\leq  \frac{m}{2\sqrt 3}$.

When the  code is self-dual the bound is stricter. In this case $\C_B^*$ itself is also a subcode of $\C$, and therefore its Hamming distance also imposes an upper bound on the spectral gap, so that in addition to the bounds discussed above, we also have $d(\C_B)\leq d(\C_B^*)$.

\subsection{Generator matrices for code CFTs}\label{sec:generating_matrix}

First we note that the lattice $3^{1/4}\tilde{\Lambda}(\C)$ is the construction A lattice obtained from a code over $F_4$, see \eqref{CA}. As a result, each vector $(3^{-1/4}\vec v,3^{1/4}\vec u) \in \tilde{\Lambda}(\C)$ must have the following property:
$u_i,v_i$ for each $i$ are simultaneously integer or half-integer.
This follows from the image of the elements $x\in F_4$ via Construction A, which maps $0,1,\omega,\omegab$ to $(u,v)=(0,0),(1,0),(-1/2, 1/2),(-1/2, -1/2)$ respectively.
Accordingly, the description of all lattices associated with codes via Construction N is as follows.  A Narain lattice with every vector $(3^{-1/4}\vec v,3^{1/4}\vec u)$ satisfying 
\begin{equation}\label{eq:evvecs}
2u_i,2v_i,v_i+u_i\in \Z
\end{equation}
can be unambiguously mapped back to an N-type code. Permutations of letters, at the level of lattices, are the T-duality transformations permuting coordinates $u_i$ and $v_i$. There are other T-duality transformations, namely orthogonal transformations $O_L \times O_R$, which commute with the metric $g_S$ and preserve the condition  \eqref{eq:evvecs}. In case of \cite{Dymarsky:2020qom} all T-dualities, at the level of codes were code equivalences. We leave open the question if this is also the case for the codes/theories of N-type. Furthermore, in the case of binary stabilizer codes, they and  corresponding theories can be parameterized by graphs, with the T-duality inducing equivalence conditions on the latter. It would be interesting to develop an analogous formalism for the codes of N-type. We take first steps in this direction below.

We show that for N-type codes, the corresponding lattices and hence code theories can be effectively parametrized 
by a handful of matrices satisfying simple constraints.  
This is essentially a generalization of the ``canonical form'' applicable both to codes and associated lattice. To derive it,  we start by reminding basic properties of the Narain lattices. Every Narain lattice $\Lambda$ which is even and self-dual with respect to the metric $g$ admits a generator matrix of the form  
\begin{equation}
 \Lambda=\twomatrix{\gamma^*}{B\gamma}{}{\gamma}
\end{equation}
where $\gamma^*=(\gamma^{-1})^T$ is the dual lattice to $\gamma$ and $B$ is an antisymmetric matrix. Now, since 
\begin{equation}
g_S=\twomatrix{\mathbb{1}_n}{}{}{S}g\twomatrix{\mathbb{1}_n}{}{}{S}\;,
\end{equation}
then if $\Lambda$ is even and self-dual with respect to $g$, then $\Lambda'=\twomatrix{\mathbb{1}_n}{}{}{S}\Lambda\twomatrix{\mathbb{1}_n}{}{}{S}$ is even and self-dual with respect to $g_S$, and vice-versa. Thus we learn  for every lattice $\tilde \Lambda (\C)$ obtained via construction \ref{constructionN}, the generator matrix can always be brought to the form
\begin{equation}
\tilde \Lambda(\C)=\twomatrix{\gamma^*}{B\gamma S}{}{S\gamma S}
\end{equation}
For convenience, we will redefine $\gamma$ and $B$ such that the generator matrix $\tilde \Lambda(\C)$ takes the form
\begin{equation}\label{eq:gen_matrix_form}
\tilde\Lambda(\C)=\twomatrix{\gamma^*}{B\gamma S}{}{\sqrt 3 S\gamma S}/3^{1/4}\;.
\end{equation}

We may now use lattice equivalences to bring $\tilde{\Lambda}(\C)$ to the simplest form possible, by bringing $\gamma^*,\gamma$ and $B$ to their ``canonical'' forms. We perform this analysis in appendix \ref{app:generating_matrix}. The result is that the generator matrix $\Lambda$ for a code theory associated with a given pairwise permutation $S$ can always be brought to the form \eqref{eq:gen_matrix_form}, where $\gamma^*,\gamma,S$ and $B$ are given as follows:
\begin{itemize}
	\item $\gamma^*$ is obtained by taking the generator matrix for the construction A lattice obtained from the binary subcode $\C_B\subset \C$ via \eqref{eq:construction_A_binary}, and further multiplying it by $\sqrt 2$. The binary subcode is an $[n,k,d_B]$ linear code, and so $\gamma^*$ can always be brought to the form 
	\begin{equation}\label{eq:general_gammastar}
	\gamma^*=\twomatrix{2\,\mathbb{1}_{n-k}}{b^T}{}{\mathbb{1}_k}
	\end{equation}
	for some $k\times (n-k)$ matrix $b$ which takes values in $\{0,1\}$ and which completely specifies the binary subcode. This means that
	\begin{equation}
		\gamma=(\gamma^{*-1})^T=\twomatrix{\frac12 \mathbb{1}_{n-k}}{}{-\frac12 b}{\mathbb{1}_{k}}\;.
	\end{equation}
	\item After putting $\gamma^*$ into this form, $S$ is not necessary in the canonical form \eqref{eq:perm_matrix}; instead, it is equivalent to the form \eqref{eq:perm_matrix} up to permutations of the rows and columns. Thus, $S$ is some matrix whose elements are $0,1$ and which obeys $S^2=1$ and $S^T=S$. We can represent $S$ as a block matrix:
	\begin{equation}\label{eq:block_S}
	S=\twomatrix{S_{11}}{S_{12}}{S_{12}^T}{S_{22}}\;,
	\end{equation}
	where $S_{11}$ is an $(n-k)\times (n-k)$ matrix, and $S_{22}$ is $k\times k$.
	\item $B$ takes the form
	\begin{equation}\label{eq:general_B}
	B=\left(\begin{array}{cc}
	\tilde B+b^TS_{12}^T-S_{12}b & b^{T}S_{22}-S_{12}\\
	S_{12}^{T}-S_{22}b & 0
	\end{array}\right)\;,
	\end{equation}
	with $\tilde B$ an integer $(n-k)\times (n-k)$ antisymmetric matrix defined mod 4.
\end{itemize}
In addition, due to the constraint \eqref{eq:evvecs}, we must have
\begin{equation}\label{eq:gen_matrix_constraint}
\tilde B+b^TS_{12}^T-S_{12}b-b^{T}S_{22}b+S_{11}=\tilde B+ \left(\mathbb{1}_{n-k}|b^T\right)S\begin{pmatrix}\mathbb{1}_{n-k}\\ b\end{pmatrix}\equiv 0 \mod\, 2\;.
\end{equation}
This is a complicated constraint in general. But if we focus on the diagonal, we find that since $\tilde B$ is antisymmetric, this reduces to the constraint that the matrix $(\mathbb{1}_{n-k}|b^T)S\begin{pmatrix}\mathbb{1}_{n-k}\\ b\end{pmatrix}$ has 0's on the diagonal. This means that the code $\C_B$ must be even with respect to the inner product $(\cdot,\cdot)_{B,S}$, which was indeed proven to be the case in section \ref{sec:binary_subcode}.

To summarize, we have found that generating matrices for code CFTs obtained via construction \ref{constructionN} are completely fixed in terms of the given permutation $S$ (i.e.~a matrix whose elements are $0,1$ and which obeys $S^2=S^T S=1$), along with a $k\times (n-k)$ binary matrix $b$ and an antisymmetric integral $(n-k)\times(n-k)$ matrix $\tilde B$ defined mod $4$ and obeying \eqref{eq:gen_matrix_constraint}. In terms of these matrices, the matrix $\gamma^*$ is given by \eqref{eq:general_gammastar}, $B$ is given by \eqref{eq:general_B}, and the full generator matrix is given by \eqref{eq:gen_matrix_form}. 

We thus have a full classification of all possible generator matrices, and so one can find all code CFTs, up to T-duality, by generating all possible building blocks $S,\gamma^*,\tilde B$ obeying these constraints. Specifically, the procedure is the following. Start by choosing a $k\times (n-k)$ matrix b which takes the values 0 or 1, which specifies $\gamma^*$ using \eqref{eq:general_gammastar}. Next, choose a permutation matrix $S$. This is an $n\times n$ symmetric matrix which obeys $S^2=S^TS=1$ and takes the values $0$ or 1, such that in every row and column exactly one element is nonzero. This defines a permutation of the bits. Finally, we must choose a matrix $\tilde B$. This is an $(n-k)\times(n-k)$ integral antisymmetric matrix defined mod 4. This matrix must also obey the condition \eqref{eq:gen_matrix_constraint}. Since we have already chosen $b$ and $S$, this constraint can be easily solved, and allows for two choices for the value of each element of $\tilde B$. Specifically, if the corresponding element of $\left(\mathbb{1}_{n-k}|b^T\right)S\begin{pmatrix}\mathbb{1}_{n-k}\\ b\end{pmatrix}$ is zero mod 2, then the element of $\tilde B$ can be 0 or 2. Otherwise, the element of  $\tilde B$ can be 1 or 3.

Finally, we discuss the simplifications that occur for the construction \ref{constructionNp}. In this case, $S=\mathbb{1}_n$ and so in equation \eqref{eq:block_S} we find $S_{11}=\mathbb{1}_k$, $S_{22}=\mathbb{1}_{n-k}$ and $S_{12}$ is the zero matrix. The main simplification is in the constraint \eqref{eq:gen_matrix_constraint}. Using the fact that the dual of the binary subcode $\C_B^*$ is self-orthogonal in this case, the constraint reduces simply to the requirement that the elements of
$\tilde B$ must be even. Thus to generate all code theories of this type we choose $b$ as above, and in addition choose an (anti)symmetric $(n-k)\times (n-k)$ matrix $\tilde B$ which is defined mod 4 and can take only even values.

\subsection{Partition functions}\label{sec:partition_functions}

The partition function of a code CFTs is related to enumerator polynomial of the corresponding code $\C$, similarly to the discussions in \cite{Dymarsky:2020qom}. However, our construction relates the partition function to an extended enumerator polynomial $W^S$ which depends on the permutation $S$. $W^S$ is defined for a pair: a code and the pairwise permutation $S$. It is a polynomial in $14$ variables:
\begin{equation}\label{eq:extended_enumerator}
W_{\C}^S(\{t_x\},\{t_{xy}\})=\sum_{c\in\C}\prod_{x\in F_4} t_x^{w_x(c)}\sideset{}{'}\prod_{x,y\in F_4} t_{xy}^{w_{xy}(c)}\;,
\end{equation}
where $\sideset{}{'}\prod_{x,y}$ means that  only ordered  pairs $(x,y)$ are included in the product. We thus have 4 variables $t_x$ and 10 variables $t_{xy}$ (because of symmetry $t_{xy}=t_{yx}$).
 $w_x(c)$ counts how many non-permuted letters $x\in F_4$ appear in  each codeword $c$, while $w_{xy}(c)$ counts how many pairs of letters $(x,y)$ are permuted into each other. 
In the case of a trivial permutation $S=1$, $w_{xy}=0$ for all codewords, and $W^S_{\C}$ reduces to the standard full enumerator polynomial $W(\{t_x\})$.

The partition function \eqref{eq:narain_partition} of a code CFT can be written in terms of $W^S$,
\begin{equation}
\label{ZCFT}
Z_{\Lambda(\C)}=\frac{1}{|\eta|^{2c}}W^S_{\C}(\{\psi_x\},\{\psi_{xy}\})\;,
\end{equation}
where the $\psi$'s are defined as follows. We define $\vec k = (k_1,k_2)\in\mathbb Z^2$, and define $g(x)$ for $x\in F_4$ to be the Gray map:
\begin{equation}
\begin{split}
g(0)=(0,0),&\quad g(1)=(1,1),\\
g(\omega)=(1,0),&\quad g(\omegab)=(0,1)\;.
\end{split}
\end{equation}
In the language of the condition \eqref{eq:eveness_condition}, we have $x=a_x \omega+b_x \bar \omega$ and
$g(x)=(a_x,b_x)$. Then 
\begin{align}
\psi_x&=\sum_{k\in\mathbb{Z}^2}\exp[ {\rm v}^T\Omega_2 {\rm v}]\label{eq:psi_i}\;,\qquad {\rm v}=2k+g(x)\;,\\
\psi_{xy}&=\sum_{k\in\mathbb{Z}^4}\exp[{\rm v}^T\Omega_4 {\rm v}]\label{eq:psi_ij}\;,\qquad {\rm v}=2k+(g(x),g(y))\;,
\end{align}
where
\begin{equation}
\Omega_2=i\pi \left(
\begin{array}{cc}
\frac{{i\tau_2}}{\sqrt{3}}-\frac{\tau_1}{2} & -\frac{{i\tau_2}}{2 \sqrt{3}} \\
-\frac{{i\tau_2}}{2 \sqrt{3}} & \frac{\tau_1}{2}+\frac{{i\tau_2}}{\sqrt{3}} \\
\end{array}
\right)\;,\quad
\Omega_4=i\pi \left(
\begin{array}{cccc}
\frac{{i\tau_2}}{\sqrt{3}} & -\frac{{i\tau_2}}{2 \sqrt{3}} & -\frac{\tau_1}{2} & 0 \\
-\frac{{i\tau_2}}{2 \sqrt{3}} & \frac{{i\tau_2}}{\sqrt{3}} & 0 & \frac{\tau_1}{2} \\
-\frac{\tau_1}{2} & 0 & \frac{{i\tau_2}}{\sqrt{3}} & -\frac{{i\tau_2}}{2 \sqrt{3}} \\
0 & \frac{\tau_1}{2} & -\frac{{i\tau_2}}{2 \sqrt{3}} & \frac{{i\tau_2}}{\sqrt{3}} \\
\end{array}
\right)\;,
\end{equation}
and we have written $\tau=\tau_1+i\tau_2$ for $\tau_1,\tau_2\in\mathbb R$.

Now we would like to verify modular invariance of the  partition function \eqref{ZCFT}.  First we discuss S-transformation $\tau\to -1/\tau$. We start with Poisson resummation of a general theta-series of the form 
\begin{equation}\label{eq:Poisson}
\sum_{\vec{n}\in \mathbb{Z}^N}e^{(2\vec n+\vec c)^{T}\Omega(2\vec n+\vec c)}=
\frac{1}{\sqrt{\det\left(-4\Omega/\pi\right)}}\sum_{\vec{m}\in \mathbb{Z}^N}e^{\frac{1}{4}\pi^{2}\vec{m}^T\Omega^{-1}\vec m+i\pi \vec{m}^Tc}\;.
\end{equation}
where $\Omega$ is an $N\times N$ matrix $\Omega$ and $c$ is an $N$-dimensional vector. 
Using \eqref{eq:Poisson} and the fact that both $\Omega_2,\Omega_4$ obey $\Omega^{-1}(\tau)=\frac{4}{\pi^2}\Omega(-1/\tau)$ up to signs which can be removed by redefining some of the integers in the sum, we  find transformations of the $\psi$'s.  To write them concisely we introduce auxiliary variables
\begin{equation}\label{eq:tx_transf}
\begin{split}
t_0'&=\frac{t_0+t_1+t_\omega+t_\omegab}{2},\\
t_1'&=\frac{t_0+t_1-t_\omega-t_\omegab}{2},\\
t_\omega'&=\frac{t_0-t_1+t_\omega-t_\omegab}{2},\\
t_\omegab'&=\frac{t_0-t_1-t_\omega+t_\omegab}{2},\\
\end{split}
\end{equation}
as well as 
\begin{equation}\label{eq:txy_transf}
\begin{split}
t_{\omega0}'	&=\frac{1}{4}(t_{00}-t_{11}-t_{\omegab\omegab}+t_{\omega \omega }+2t_{\omega 0}-2t_{\omegab1}),\\
t_{\omegab 0}'&=\frac{1}{4}(t_{00}-t_{11}+t_{\omegab\omegab}-t_{\omega \omega }+2t_{\omegab0}-2t_{\omega 1})\\
t_{10}'	&=\frac{1}{4}(t_{00}+t_{11}-t_{\omegab\omegab}-t_{\omega \omega }+2t_{10}-2t_{\omegab\omega }),\\
t_{\omegab \omega}'	&=\frac{1}{4}(t_{00}+t_{11}-t_{\omegab\omegab}-t_{\omega \omega }-2t_{10}+2t_{\omega\omegab }),\\
t_{1\omega}'	&=\frac{1}{4}(t_{00}-t_{11}+t_{\omegab\omegab}-t_{\omega \omega }-2t_{\omegab0}+2t_{\omega 1}),\\
t_{1\omegab }'&=\frac{1}{4}(t_{00}-t_{11}-t_{\omegab\omegab}+t_{\omega \omega }-2t_{\omega 0}+2t_{\omegab1}),\\
t_{00}'	&=\frac{1}{4}(t_{00}+t_{11}+t_{\omegab\omegab}+t_{\omega \omega }+2t_{\omegab0}+2t_{\omega 0}+2t_{10}+2t_{\omegab1}+2t_{\omega 1}+2t_{\omegab\omega }),\\
t_{11}'	&=\frac{1}{4}(t_{00}+t_{11}+t_{\omegab\omegab}+t_{\omega \omega }-2t_{\omegab0}-2t_{\omega 0}+2t_{10}-2t_{\omegab1}-2t_{\omega 1}+2t_{\omegab\omega }),\\
t_{\omega\omega}'	&=\frac{1}{4}(t_{00}+t_{11}+t_{\omegab\omegab}+t_{\omega \omega }+2t_{\omegab0}+2t_{\omega 0}-2t_{10}+2t_{\omegab1}-2t_{\omega 1}-2t_{\omegab\omega }),\\
t_{\omegab \omegab }'	&=\frac{1}{4}(t_{00}+t_{11}+t_{\omegab\omegab}+t_{\omega \omega }+2t_{\omegab0}+2t_{\omega 0}-2t_{10}-2t_{\omegab1}+2t_{\omega 1}-2t_{\omegab\omega }).
\end{split}
\end{equation}
In terms of these variables the analog of the MacWilliams identity takes the form
\begin{equation}
W^S_{\C^*}(\{t_x\},\{t_{xy}\})=
W^S_{\C}(\{t'_{x}\},\{t'_{xy}\})\;.
\end{equation}
Note that for $S=1$ this reduces to the standard MacWilliams identity for codes over $F_4$ 
\cite{nebe2006self}. For the N-type codes satisfying $\C^*=S(\bar \C)$ this  yields
\begin{equation}
W^S_{\C}(\{t_x\},\{t_{xy}\})=
W^S_{\C}(\{t'_{\bar x}\},\{t'_{\bar x\bar y}\})\;.
\label{eq:mac}
\end{equation}

We can now write down the transformation of the $\psi$'s under $\tau \rightarrow -1/\tau$. Transformations of $\psi_x$ are exactly the same as the transformation $t_x\to t_x'$ in \eqref{eq:tx_transf}, supplemented by the conjugation  $\omega \leftrightarrow \bar \omega$, i.e.~the interchange of $t_\omega$ with $t_{\bar \omega}$ etc.~in the RHS of \eqref{eq:tx_transf}, and also multiplied by $\sqrt{\tau\bar{ \tau}}$. For example, compare with \eqref{eq:tx_transf},
\begin{equation}
    \begin{split}
    \psi_\omega(-1/\tau)&=\sqrt{\tau\bar{\tau}}\frac{\psi_0(\tau)-\psi_1(\tau)-\psi_\omega(\tau)+\psi_\omegab(\tau)}{2}.
    \end{split}
\end{equation}
Similarly, $\psi_{xy}$ transforms just like $t_{xy}$ in \eqref{eq:txy_transf}, supplemented by the interchange  of $\omega,\omegab$ on the RHS and multiplied by $\tau\overline{\tau}$. For example, compare with \eqref{eq:txy_transf},
\begin{equation}
\begin{split}
\psi_{\omega0}\left(-1/\tau\right)	&=\frac{\tau\overline{\tau}}{4}(\psi_{00}-\psi_{11}-\psi_{\omega\omega}+\psi_{\omegab \omegab }+2\psi_{\omegab 0}-2\psi_{\omega1}),\\
\psi_{\omegab \omegab }\left(-1/\tau\right)	&=\frac{\tau\overline{\tau}}{4}(\psi_{00}+\psi_{11}+\psi_{\omega\omega}+\psi_{\omegab \omegab }+2\psi_{\omega0}+2\psi_{\omegab 0}-2\psi_{10}-2\psi_{\omega1}+2\psi_{\omegab 1}-2\psi_{\omega\overline{\omega}}).
\end{split}
\end{equation}
Due to these transformations, we can immediately check that the partition function is invariant under $S$-duality. Using the Macwilliams identity \eqref{eq:mac}, we find
\begin{equation}
    W^S_{\C}({(\tau\bar\tau)^{-1/2}}\psi_{x}(-1/\tau)\},\{(\tau\bar\tau)^{-1}\psi_{xy}(-1/\tau)\})=W^S_{\overline{\C}^*}(\{\psi_{x}(\tau)\},\{\psi_{xy}(\tau)\})\;.
\end{equation}
Next, using the fact that $W^S_{\C}=W^S_{S(\C)}$ (since  a permutation does not change any of the weights $w_{x},w_{xy}$),
and $C^*=S(\overline{C})$ for N-type codes, we find 
\begin{equation}\label{eq:S_transf}
    W^S_{\C}(\{\psi_{x}(-1/\tau)\},\{\psi_{xy}(-1/\tau)\})=(\tau\bar\tau)^{n/2} W^S_{\C}(\{\psi_{x}(\tau)\},\{\psi_{xy}(\tau)\})\;.
\end{equation}
This ensures invariance of the  CFT partition function \eqref{ZCFT} under the S-transformation.

Next we discuss  T-transformation $\tau\to\tau+1$. Looking at $\Omega_2,\Omega_4$, we find it acts  on $\psi_x$ by introducing a phase, $\psi_x\to \exp(\frac{i\pi}{2} (b^2-a^2))\psi_x$ where $g(x)=(a,b)$, and similarly  $\psi_{xy}\to \exp(i\pi (b_xb_y-a_xa_y))\psi_{xy}$. Thus, the contribution from every codeword $c$ corresponds to a phase $\exp(\frac{i\pi}{2}\sum_{i}(b_{S(i)}b_i-a_{S(i)}a_i) )$. Due to the evenness condition \eqref{eq:eveness_condition}, this phase is always 1, and so the partition function is invariant under the T-transformation.
This corresponds to the following symmetry of $W^S$:
\begin{equation}
\label{eq:T_transf}
    W^S_{\C}(\{{ t}_x\},\{{ t}_{xy}\})=W^S_{\C}(\{{\tilde t}_x\},\{{\tilde t}_{xy}\})\;,
\end{equation}
where
\begin{equation}
\begin{split}
\tilde{t}_\omega&= -i t_\omega,\quad \tilde{t}_\omegab= i t_\omegab\\
\tilde{t}_{1\omega}&=  -t_{1\omega},\quad
\tilde{t}_{1\omegab}= -t_{1\omegab},\\
\tilde{t}_{\omega\omega}&=  -t_{\omega\omega},\quad
\tilde{t}_{\omegab\omegab}= -t_{\omegab\omegab}\;.
\end{split}
\end{equation}
and all other $t$'s are invariant.

\subsubsection{Classification of Enumerator Polynomials}
\label{sec:invpol}
We now attempt to classify the extended enumerator polynomials of the codes of N-type. In fact, our task is broader, it is to  classify all homogeneous polynomials of 14 variables which are invariant under the S-transformation \eqref{eq:mac} and the T-transformation \eqref{eq:T_transf}. In general, the two transformations generate a group $G$ acting on the polynomials of  14 variables, and so we are looking for the ring of polynomials invariant under $G$. One can write down a generating function describing the dimension of the space of invariant homogeneous polynomials of degree $n$, called the Molien series 
\begin{equation}
	M(r)=\sum_{n=0}^\infty\dim(R_n^G)\, r^n\;.
\end{equation}
Here $R_n^G$ is the space of all polynomials of degree $n$ invariant under $G$. Molien's formula (see e.g. \cite{mukai_2003})  gives a simple expression for this generating function:
\begin{equation}
M(r)=\frac{1}{|G|}\sum_{g\in G}\frac{1}{\det(1-rg)}\;.
\end{equation}
Here, $|G|$ is the number of elements in $G$, and $g$ is a $14\times14$ matrix acting on the individual  variables.

We start by classifying polynomials associated with codes of the simpler construction \ref{constructionNp}. In this case there is no permutation, and so $W^S$ reduces to the standard enumerator polynomial of 4 variables, 
\begin{equation}
W(t_0,t_1,t_{\omega},t_{\omegab})\;.
\end{equation}
The symmetry group $G$ is generated by two matrices acting on the vector $(t_0,t_1,t_{\omega},t_{\omegab})$ as follows
\begin{equation}
S=\frac{1}{2}\left(
\begin{array}{cccc}
1 & 1 & 1 & 1 \\
1 & 1 & -1 & -1 \\
1 & -1 & 1 & -1 \\
1 & -1 & -1 & 1 \\
\end{array}
\right),\;\;\;\;\; 
T=\left(
\begin{array}{cccc}
1 & 0 & 0 & 0 \\
0 & 1 & 0 & 0 \\
0 & 0 & i & 0 \\
0 & 0 & 0 & -i \\
\end{array}
\right)\;.
\end{equation}
These matrices obey $S^2=1, T^4=1$ and $(ST)^6=1$, and so the group generated by all possible products of $T,S$ is finite and consists of 48 elements. The Molien series is 
\begin{equation}
M(r)=\frac{1}{(r-1)^4 (r+1)^3 \left(r^6+2 r^4+2 r^2+1\right)}=1+r+2r^2+2r^3+4r^4+4r^5+7r^6+7r^7+...
\end{equation}

The Molien series gives the total dimension of the space of invariant polynomials of degree $n$, yet we are interested in 
finding the generators of the polynomial ring invariant under $G$, i.e.~those polynomials which cannot be written as a product of the lower-order ones. We denote by $m_n'$ the the dimension of the space of invariant polynomials of degree $n$ which cannot be written as sums and products of lower-order invariant polynomials. Then $m_n'$ can be obtained using a recursion formula:
\begin{equation}
m_k'=m_k-\sum_{p\in P_k,p\neq (k)}\prod_{i=1}^{k-1}\begin{pmatrix}
w_i(p)+m_i'-1\\w_i(p)
\end{pmatrix}\;,\qquad m_k=\dim(R_k^G)\; .
\end{equation}
Here $P_k$ denotes all integer partitions of $k$, and we are summing over all partitions apart from the trivial partition $(k)$. In addition, $w_i(p)$ counts how many times the number $i$ appears in the partition $p$.

In the case at hand the nonzero values of $m'_k$ are $m_1',m_2',m_4',m_6'$, which are all equal to 1. This means the ring of invariant polynomials is generated by four polynomials of dimensions $1,2,4,6$. We can find them explicitly,
\begin{equation}\label{eq:invar_pols}
\begin{split}
p_1=&t_0+t_1\;,\\
p_2=&t_0^2+t_1^2+2t_{\omega}t_{\omegab}\;,\\
p_4=&t_0^4+6 t_0^2 t_1^2+t_1^4+t_{\omega}^4+6 t_{\omega}^2 t_{\omegab}^2+t_{\omegab}^4\;,\\
p_6=&t_0^6+24 t_0 t_1 t_{\bar{\omega }}^2 t_{\omega }^2+3 t_0^4 t_1^2+8 t_0^3 t_1^3+3 t_0^2 t_1^4+6 t_0^2 t_{\bar{\omega }}^2 t_{\omega }^2\\
&+3 t_0^2 t_{\bar{\omega }}^4+3 t_0^2 t_{\omega }^4+6 t_1^2 t_{\bar{\omega }}^2 t_{\omega }^2+3 t_1^2 t_{\bar{\omega }}^4+3 t_1^2 t_{\omega }^4+t_1^6\;.
\end{split}
\end{equation}
To conclude, any full enumerator polynomial of a code over $F_4$ which obeys the conditions outlined in construction \ref{constructionNp} can be written in terms of the four polynomials above. In particular, each $p_i$ corresponds to a specific N-type code, and we explicitly describe the codes for $i=1,2,4$ in section \ref{sec:simple_examples}. The code for $i=6$ is also known, but not discussed in this paper. 

We can now discuss the more general case of isodual N-type codes, for which permutation $S$ is non-trivial. The symmetry group is again generated by two matrices $T,S$, which are now $14\times 14$ matrices, and which again obey $T^4=S^2=(ST)^6=1$. The group again consists of 48 elements, but the Molien series in this case is more complicated:
\begin{equation}
\begin{split}
    M(r)=&\frac{1}{(r-1)^{14} (r+1)^6 \left(r^2+1\right) \left(r^2+r+1\right)^4}\left(1+3 r^2+12 r^3+26 r^4+30 r^5+56 r^6+58 r^7\right.\\
    &\left.+60 r^8+58 r^9+56 r^{10}+30 r^{11}+26 r^{12}+12 r^{13}+3 r^{14}+r^{16}\right)\;.
\end{split}
\end{equation}
The values of $m_k'$ for small $k$ are
\begin{equation}
m_k'=\{4,8,16,21,-6,...\}\;.
\end{equation}
We emphasize that $k$ here is not the length of the code, but is the order of the polynomial in terms of the variables $t_x,t_{xy}$. For example, at $k=1$ this includes the polynomial $t_{00}$, which corresponds to a code with $n=2$. 

There are two things to note here. First, even at small $k$ there are many generators of this ring. Second and more surprising, there are now also \textit{negative} values of $m_k'$. This means the generators of the ring are not independent, and in fact there are relations between them. To the extent of our knowledge, this is the first example when the ring of invariant polynomials associated with the class of codes is not a freely generated one, but involves non-trivial relations between generators. The first relation occurs at order $k=5$ and involves many dozens of polynomials of smaller degree. It is way too cumbersome to be written down here. 

To conclude the discussion, 
 we write all independent enumerator polynomials for $n=2$ codes (of course the choice of independent polynomials is ambiguous). Without permutation, i.e.~with $S=1$, there is only one such polynomial, which is just $p_2$ defined above. Next, we consider the permutation of the two bits. The corresponding polynomial is of degree one, i.e. it has $k=1$, and so we must look at $m'_1$ which is equal to $4$. This includes all possible polynomials of degree one, including those without a permutation, and so we must subtract polynomials without a permutation, of which there is one -- $p_1$ defined above. Thus, there should be three independent invariant polynomials of degree $k=1$ involving a permutation of two letters:
\begin{equation}
\begin{split}
q_1=&t_{00}+t_{11}+2t_{01}\;,\\
q_2=&t_{00}+t_{11}+2t_{\omega\omegab}\;,\\
q_3=&t_{00}+t_{0\omega}+t_{0\omegab}+t_{\omega\omegab}\;.
\end{split}
\end{equation}

\section{Simple examples}\label{sec:simple_examples}

In this section we describe a number of examples of using the constructions \ref{constructionN} and  \ref{constructionNp}, including an explicit form for the generator matrices for the corresponding Narain CFTs.

\subsection{Binary codes over $F_4$}\label{sec:binary}

Define the code $\mathcal{B}_n=F_2^n$ over $F_4$, i.e.~$\mathcal{B}_n$ includes all possible codewords whose elements are either $0$ or $1$ (in particular, $\mathcal{B}_n$ is equal to its own binary subcode). Its enumerator polynomial is
\begin{eqnarray}
p_1^n=(t_0+t_1)^n\;.
\end{eqnarray}
This code is an N-type code for any choice of permutation $S$, and so for any $S$ we can use Construction \ref{constructionN} to generate a Narain CFT from it. We note that for any $n$, $\mathcal{B}_n$ is the unique N-type code whose letters are all either $0$ or $1$.\footnote{To see this, note that any N-type code must have $2^{n}$ codewords, and the only way to generate this number using binary codewords is to include all possible binary codewords.}

The lattice $\tilde{\Lambda}(\mathcal{B}_n)$ obtained via construction \ref{constructionN} 
includes all $n$-dimensional vectors of the form $(3^{-1/4}\vec v,3^{1/4}\vec u)$ where  $v_i, u_i\in\mathbb{Z}$. We can construct its generator matrix explicitly following section \ref{sec:generating_matrix}. There we demonstrated that the generator matrix can be brought to the form 
\begin{equation}
\tilde\Lambda(\mathcal{B}_n)=\twomatrix{\gamma^*}{B\gamma S}{}{\sqrt 3 S\gamma S}/3^{1/4}\;,
\end{equation}
with $\gamma^*$ the generator matrix of the construction A lattice of the binary subcode of $\mathcal{B}_n$ multiplied by an additional factor $\sqrt 2$. In our case, 
\begin{equation}
\gamma^*=\gamma=\mathbb{1}_n,
\end{equation}
and we can use T-dualities to set $B=0$ by adding columns of $\begin{pmatrix}
\gamma^*\\0
\end{pmatrix}$ with $\gamma^*=\mathbb{1}_n$ to $\begin{pmatrix}
B \gamma S\\\sqrt3 S\gamma S
\end{pmatrix}$. The generator matrix is thus especially simple and takes the form
\begin{equation}
\tilde\Lambda(\mathcal{B}_n)=\twomatrix{\mathbb{1}_n}{}{}{\sqrt 3\, \mathbb{1}_n}/3^{1/4}\;.
\end{equation}
The spectral gap is always $\Delta_1=\frac{1}{2\sqrt 3}$.

\subsection{Codes with $n=1$}

We now describe all N-type codes with $n=1$. There is only one such code, which is the binary code discussed in \ref{sec:binary}:
\begin{equation}
    \mathcal{B}_1=\{ (0),(1) \}\;,
\end{equation}
with enumerator polynomial $p_1=t_0+t_1$ and the generator matrix, up to T-dualities, is
\begin{equation}
\tilde\Lambda(\mathcal{B}_1)=\twomatrix{3^{1/4}}{}{}{3^{-1/4}}\;.
\end{equation}
Corresponding theory is a compact boson at radius $R=\sqrt{2}\,3^{1/4}$, which is a non-rational theory.

\subsection{Codes with $n=2$}

We now describe all N-type codes with $n=2$ up to the automorphisms discussed in section \ref{sec:partition_functions}. At $n=2$, there are two options for pairwise permutations: no permutation and the permutation interchanging the two letters. We discuss each one separately.

\subsubsection{No permutation}
\label{531}
If there is no permutation, then there are two N-type codes (and for both we can apply the simple construction \ref{constructionNp} instead of the general one). One is the binary code $\mathcal{B}_2$ defined in \ref{sec:binary}, with enumerator polynomial $p_1^2=(t_0+t_1)^2$. The other code is
\begin{equation}
\C_2 =\{ (0,0),(1,1),(\omega,\overline\omega),(\overline\omega,\omega) \}\;.
\end{equation}
It has enumerator polynomial 
\begin{eqnarray}
p_2=t_0^2+t_1^2+2t_{\omega}t_{\omegab}\;.
\end{eqnarray}
This code is self-dual and has $w_\omega(c)=w_{\overline\omega}(c)$ for every codeword $c$, and so we can apply construction \ref{constructionNp} to construct a Narain lattice $\tilde{\Lambda}(\C_2)$ from it. 
The generator matrix takes the form
\begin{equation}
\tilde\Lambda(\C_2)=\twomatrix{\gamma^*}{B\gamma }{}{\sqrt 3 \gamma }/3^{1/4}\;,
\end{equation}
where $\gamma^*$ is the generator matrix of the construction A lattice of the binary subcode of $\C_2$ times $\sqrt 2$, $\gamma$ is given by $\gamma=(\gamma^{*-1})^T$, and $B$ takes the form outlined in the equation \eqref{eq:general_B} with $S=1$. We now construct this generator matrix explicitly. 

First we find $\gamma^*$. The binary subcode of $\C_2$ is $\{ (0,0),(1,1) \}$, and so $\gamma^*$ can be brought to the form
\begin{equation}
\gamma^*=\twomatrix{2}{1}{}{1}\;. 
\end{equation}
In particular, comparing $\gamma^*$ to the canonical form \eqref{eq:general_gammastar}, we find that $b=1$. Next we study the matrix $B$. Following equation \eqref{eq:general_B}, it must take the form
\begin{equation}
B=\twomatrix{\tilde B}{b^T}{-b}{}\;,
\end{equation}
with $\tilde B$ being an antisymmetric $1\times 1$ matrix. This means $\tilde B=0$, and so the final form of the generator matrix is
\begin{equation}
\tilde\Lambda(\C_2)=\left(
\begin{array}{cccc}
2 & 1 & -\frac{1}{2} & 1 \\
0 & 1 & -\frac{1}{2} & 0 \\
0 & 0 & \frac{\sqrt{3}}{2} & 0 \\
0 & 0 & -\frac{\sqrt{3}}{2} & \sqrt{3} \\
\end{array}
\right)/3^{1/4}\;.
\end{equation}
The spectral gap of this CFT is $\Delta_1=\frac{1}{\sqrt3}$.

\subsubsection{With a permutation}

If we include a permutation of the two letters,  there are four N-type codes up to automorphisms:
\begin{equation}
\label{532}
    \begin{split}
        \mathcal{B}_2&=\{ (0,0),(0,1),(1,0),(1,1) \}\;,\\
        \C_2&=\{ (0,0),(1,1),(\omega,\overline\omega),(\overline\omega,\omega) \}\;,\\
        \tilde{\C}_2&=\{ (0,0),(0,\omega),(\omegab,0),(\omegab,\omega) \}\;,\\
        \mathcal{C}'_2&=\{ (0,0),(0,1),(0,\omega),(0,\omegab) \}\;,
    \end{split}
\end{equation}
Their extended enumerator polynomials are
\begin{equation}
\begin{split}
W^S(\mathcal{B}_2)&=q_1=t_{00}+2t_{01}+t_{11}\;,\\
W^S(\mathcal{C}_2)&=q_2=t_{00}+t_{11}+2t_{\omega\omegab}\;,\\
W^S(\tilde{\mathcal{C}}_2)&=q_3=t_{00}+t_{0\omega}+t_{0\omegab}+t_{\omega\omegab}\;,\\
W^S(\mathcal{C}'_2)&=\frac{1}{2}q_1-\frac{1}{2}q_2+q_3=t_{00}+t_{01}+t_{0\omega}+t_{0\omegab}\;.
\end{split}
\end{equation}
We can find the generator matrices of the corresponding Narain lattices:
\begin{equation}
    \begin{split}
    \Lambda(\mathcal{B}_2)=\begin{pmatrix}
    1 & & & \\
     & 1 & & \\
     & & \sqrt 3 & \\
    & & & \sqrt 3 \\
    \end{pmatrix}/3^{1/4},&\quad
    \Lambda(\C_2)= 
    \begin{pmatrix}
     2 & 1 & 1 & -\frac{1}{2} \\
      & 1 &  & -\frac{1}{2} \\
      &  & \sqrt{3} & -\frac{\sqrt{3}}{2} \\
      &  &  & \frac{\sqrt{3}}{2} \\
    \end{pmatrix} /3^{1/4}   \\
    \Lambda(\tilde{\C}_2)=
    \begin{pmatrix}
     2 &  & \frac{1}{2} &  \\
      & 2 &  & -\frac{1}{2} \\
      &  & \frac{\sqrt{3}}{2} &  \\
      &  &  & \frac{\sqrt{3}}{2} \\
    \end{pmatrix}    /3^{1/4},&\quad
    \Lambda(\mathcal{C}'_2)=
\begin{pmatrix}
 2 &  & 1 &  \\
  & 1 &  & -\frac{1}{2} \\
  &  & \sqrt{3} & \\
  &  &  & \frac{\sqrt{3}}{2} \\
\end{pmatrix}
/3^{1/4}\;.\\
    \end{split}
\end{equation}
The spectral gaps of these CFTs are respectively
\begin{equation}
    \begin{split}
       \Delta_1(\mathcal{B}_2)=\frac{1}{2\sqrt3}\;,&\qquad  \Delta_1(\mathcal{C}_2)=\frac{1}{\sqrt3}\;,\\
       \Delta_1(\tilde{\mathcal{C}}_2)=\frac{1}{2\sqrt3}\;,&\qquad
       \Delta_1(\mathcal{C}_2')=\frac{1}{2\sqrt3}\;.
    \end{split}
\end{equation}
We would like to note that $\mathcal{B}_2$ in \eqref{532} is exactly the same as in the section \ref{531}, but its extended enumerator polynomial is different: $p_1^2$ and $q_1$. This is because this code is invariant under permutations and, ammended with different $S$, gives rise to different CFTs. 

\subsection{An $n=4$ code}

We now describe an N-type code with $n=4$ for which the simple construction \ref{constructionNp} applies. The code is 
\begin{equation}
   \C_4= \Bigg\{
\begin{array}{cccc}
    (0 , 0 , 0 , 0)&\;    (1 , 1 , 0 , 0)&\; (1 , 0 , 1 , 0) &\;  (1 , 0 , 0 , 1) \\
   (0 , 1 , 0 , 1)&\; (0 , 1 , 1 , 0)&\; (0 , 0 , 1 , 1)&\; (1 , 1 , 1 , 1) \\
 (\omegab  , \omegab  , \omega  , \omega)&\;
 (\omega  , \omegab  , \omegab  , \omega)&\;
 (\omega  , \omega  , \omegab  , \omegab)&\;
 (\omegab  , \omega  , \omegab  , \omega)  \\
 (\omegab  , \omega  , \omega  , \omegab)&\;
 (\omega  , \omegab  , \omega  , \omegab)&\;
  (\omega  , \omega  , \omega  , \omega)&\;
  (\omegab  , \omegab  , \omegab  , \omegab)
\end{array}
\Bigg\}\;.
\end{equation}
Its enumerator polynomial is $p_4$ from \eqref{eq:invar_pols}. This code is self-dual and has $w_\omega(c)=w_{\overline\omega}(c)$ for every codeword $c$. Applying construction \ref{constructionNp}, we get a Narain lattice $\tilde\Lambda(\C_4)$. Its generator matrix takes the form
\begin{equation}
\tilde\Lambda(\C_4)=\twomatrix{\gamma^*}{B\gamma }{}{\sqrt 3 \gamma }/3^{1/4}\;,
\end{equation}
where 
\begin{equation}
    \gamma^*=\left(
    \begin{array}{cccc}
     2 & 1 & 1 & 1 \\
     0 & 1 & 0 & 0 \\
     0 & 0 & 1 & 0 \\
     0 & 0 & 0 & 1 \\
    \end{array}
    \right),\quad B=\left(
    \begin{array}{cccc}
     0 &1 & 1 & 1 \\
     -1 & 0 & 0 & 0 \\
     -1 & 0 & 0 & 0 \\
     -1 & 0 & 0 & 0 \\
    \end{array}
    \right)\;.
\end{equation}
The spectral gap is $\Delta_1=\frac{1}{\sqrt 3}$.

\section{Constructing optimal CFTs}\label{sec:constructing_optimal}

In this section we construct the  hypothetical optimal Narain CFTs with the central charges $c=6$ and $c=7$. Both of these  were found in \cite{Afkhami-Jeddi:2020ezh}. For $c=6$, there is an obvious guess for the code associated with it. The corresponding Narain lattice is the Coxeter-Todd lattice understood as a Lorentizan lattice; the latter is known to be related to the hexacode -- the unique self-dual $[n,k,d]=[6,3,4]$ code over $F_4$. For $c=7$ the code we find is less well-known, and we dub it the ``septacode.'' Finally, we discuss optimal CFTs at other values of central charge and their possible  relations to codes.

\subsection{$c=6$ and the hexacode}
\label{sec:hexacode}

 Consider the hexacode $\mathcal{H}$, which is a linear\footnote{Linear means code generator matrix is multiplied by the elements of $F_4$. For an additive code over $F_4$ it is $F_2$.} $\left[6,3,4\right]$ code over $F_4$. We can choose the generating matrix to be \cite{SPLAG}
\begin{equation}\label{eq:hexacode_generator}
G=\left(
\begin{array}{ccc}
0 & 0 & 1 \\
0 & 1 & 0 \\
1 & 0 & 0 \\
1 & 1 & 1 \\
1 & \omega & \omegab   \\
1 & \omegab   & \omega \\
\end{array}
\right)\;.
\end{equation}
The hexacode is even and self-dual. Its Hamming distance is $d(\mathcal{H})=4$, and its binary subcode is
\begin{eqnarray}
\mathcal{H}_B=\{ (0^6),(1^2,0^2,1^2),(0^2,1^4),(1^4,0^2) \}\;.
\end{eqnarray}
The corresponding construction A lattice $\Lambda(\mathcal{H})$ is the Coxeter-Todd lattice $K_{12}$ \cite{SPLAG}.

We now show that we can apply construction \ref{constructionN} to the hexacode with a permutation $S$ that permutes the last two letters, so that the lattice $\tilde \Lambda(\mathcal{H})=K_{12}/3^{1/4}$ is a Narain lattice. The extended enumerator polynomial of the hexacode with respect to this permutation is
\begin{equation}
    \begin{split}
        W^S_{\mathcal{H}}=&t_0^4 t_{00}+t_{00} t_{\omega }^4+t_{00} t_{\omegab }^4+t_1^4 t_{00}+8 t_1 t_0 t_{01} t_{\omega } t_{\omegab }+8 t_1 t_0 t_{0\omega } t_{\omega } t_{\omegab }+8 t_1 t_0 t_{0\omegab} t_{\omega } t_{\omegab }+2 t_{11} t_{\omega }^2 t_{\omegab }^2\\
        &+2 t_1^2 t_0^2 t_{11}+4 t_1^2 t_{1\omega} t_{\omega }^2+4 t_0^2 t_{1\omega } t_{\omegab }^2+4 t_0^2 t_{1\omegab} t_{\omega }^2+4 t_1^2 t_{1\omegab} t_{\omegab }^2+4 t_{\omega }^2 t_{\omegab }^2 t_{\omega \omegab}+2 t_1^2 t_{\omega }^2 t_{\omegab\omegab}+2 t_0^2 t_{\omega }^2 t_{\omega \omega }\\
        &+2 t_0^2 t_{\omegab }^2 t_{\omegab\omegab}+2 t_1^2 t_{\omegab }^2 t_{\omega \omega }+4 t_1^2 t_0^2 t_{\omega \omegab}\;.
    \end{split}
\end{equation}

The hexacode obtained using the generating matrix  \eqref{eq:hexacode_generator} obeys $\mathcal{H}=S(\overline {\mathcal{H}})$, where $S$ interchanges the last two letters, and so using \ref{claim1} we find that $\tilde \Lambda(\mathcal{H})$ is self-dual with respect to the metric $g_S$ \eqref{eq:metric_gS}. One can check explicitly that the hexacode obeys \ref{claim2} conditions with the same permutation $S$, and so $\tilde \Lambda(\mathcal{H})$ is also even with respect to the metric $g_S$. $\mathcal{H}$ is thus an N-type code, and so using construction \ref{constructionN} we find that $\tilde \Lambda(\mathcal{H})$ is a Narain lattice. The corresponding spectral gap is $\Delta_1=\sqrt{4/3}$, which was conjectured in \cite{Afkhami-Jeddi:2020ezh} to be the maximal value for Narain CFTs with $c=6$. Indeed, the Narain lattice discussed there is just the Coxeter-Todd lattice rescaled by $3^{1/4}$.
We thus have reproduced the optimal Narain CFT at $c=6$ using construction \ref{constructionN}.

Explicitly, the generator matrix of the Narain lattice can be brought to the form \eqref{eq:gen_matrix_form} with 
\begin{equation}
\gamma^*=\twomatrix{2\,\mathbb{1}_4}{b^T}{}{\mathbb{1}_2},\;\;\;\;\; 
B=\left(\begin{array}{cc}
\tilde{B} & b^T\\
-b & 0
\end{array}\right),
\end{equation}
where
\begin{equation}
b=
\begin{pmatrix}
1&1&0&1\\
1&1&1&0
\end{pmatrix}\;,
\;\;\;\;\;
\tilde{B}=\left(
\begin{array}{cccc}
0 & -1 & 1 & -1 \\
1 & 0 & -1 & 1 \\
-1 & 1 & 0 & 2 \\
1 & -1 & -2 & 0 \\
\end{array}
\right)\;,
\end{equation}
and where the permutation matrix $S$ is different now, and is given by
\begin{equation}
S=\left(
\begin{array}{cccccc}
1 & 0 & 0 & 0 & 0 & 0 \\
0 & 1 & 0 & 0 & 0 & 0 \\
0 & 0 & 1 & 0 & 0 & 0 \\
0 & 0 & 0 & 0 & 1 & 0 \\
0 & 0 & 0 & 1 & 0 & 0 \\
0 & 0 & 0 & 0 & 0 & 1 \\
\end{array}
\right)\;,
\end{equation}
i.e.~it interchanges the fourth and fifth letters.

We note that representation of the lattice above is different from that one in \cite{Afkhami-Jeddi:2020ezh}, but must be related to it by a $O(6)\times  O(6)$ transformation.

\subsection{$c=7$ and the septacode}
\label{sec:septacode}

The conjectured optimal Narain CFT with $c=7$ has spectral gap $\Delta_1=\sqrt{4/3}$ \cite{Afkhami-Jeddi:2020ezh}.
We note there are no $n=7$ self-dual codes over $F_4$ with $d=4$, and that the maximal Hamming distance in this case is $d=3$ \cite{calderbank1998quantum}. 
Should we want to use our construction to obtain optimal $c=7$ CFT from a code, we must therefore consider isodual codes, with $\C^*=S(\overline{C})$. It turns out  this only requires $\C^*=\overline{C}$, i.e.~one can take $S=1$.

Consider the $n=7$ code $\C_7$ generated over $F_2$ by the matrix
\begin{equation}
G=\left(
\begin{array}{ccccccc}
1 & \omega  & 0 & 0 & 0 & 1 & 0 \\
1 & 1 & \omega  & 0 & 1 & 0 & 0 \\
1 & 0 & 0 & \omegab & 1 & 1 & 1 \\
1 & 1 & 0 & 1 & \omega  & 0 & 1 \\
1 & 0 & 1 & 1 & 1 & \omega  & 0 \\
1 & 1 & 1 & 1 & 0 & 1 & \omega  \\
1 & \omegab & \omegab & \omega  & \omegab & \omegab & \omegab \\
\end{array}
\right)
\end{equation}
We will call this code the "Septacode". The septacode is additive but not linear, it has $2^7$ elements and its Hamming distance is $d=4$. Its enumerator polynomial is
\begin{equation}
\begin{split}
    W_{\C_7}=&t_0^7+21 t_1^2 t_0^3 t_{\omega } t_{\omegab}+21 t_1 t_0^2 t_{\omega }^2 t_{\omegab}^2+21 t_1^3 t_0^2 t_{\omega } t_{\omegab}+7 t_0 t_{\omega }^3 t_{\omegab}^3\\
    &+21 t_1^2 t_0 t_{\omega }^2 t_{\omegab}^2+7 t_1 t_{\omega }^3 t_{\omegab}^3+7 t_0^3 t_{\omega }^4+7 t_1^3 t_{\omega }^4+7 t_0^3 t_{\omegab}^4+7 t_1^3 t_{\omegab}^4+t_1^7\;.
    \end{split}
\end{equation}
Its binary subcode is
\begin{equation}
\C_B = \{ (0^7),\;(1^7) \}\subset \C_7\;,
\end{equation}
so $\C_B$ is just the $n=7$ binary repetition code. In addition, the septacode obeys
\begin{equation}
\C_7^* = \overline{\C_7}\;.
\end{equation}
and so in particular it obeys \ref{claim1} with a trivial permutation $S=1$. It is also simple to check it obeys the \ref{claim2} conditions. It is thus an N-type code, and we can use construction 
\ref{constructionN} to obtain a Narain lattice $\tilde{\Lambda}(\C_7)$ from it.

Let us construct the generator matrix for the Narain lattice explicitly, following section \ref{sec:generating_matrix}. The generator matrix is given by the general form \eqref{eq:gen_matrix_form} with $S=1$.
$\gamma^*$ is the generator matrix for the construction A lattice of the binary subcode $\C_B\subset\C$ (times $\sqrt 2$), so in the case at hand it is
\begin{equation}
\gamma^*=2\mathbb{Z}^{7}\cup\left(2\mathbb{Z}^{7}+\left(1,...,1\right)\right)=2D_{7}^{*}\;.
\end{equation}
We can bring this to the standard form of
\begin{equation}
\gamma^*=\Lambda(\C_B)=\left(\begin{array}{cc}
2\,\mathbb{1}_{6} & b^T\\
& 1
\end{array}\right)\;, \qquad b=(1^6)\;.
\end{equation}
As discussed above, $B$ can be brought to the form 
\begin{equation}
B=\left(\begin{array}{cc}
\tilde{B} & b^T\\
-b & 0
\end{array}\right)\, ,
\end{equation}
with $\tilde{B}$ being some integral antisymmetric matrix. In the case at hand we find
\begin{equation}
\tilde B= \left(
\begin{array}{cccccc}
0 & 1 & -1 & 1 & -1 & 1 \\
-1 & 0 & -1 & -1 & 1 & 1 \\
1 & 1 & 0 & -1 & -1 & -1 \\
-1 & 1 & 1 & 0 & 1 & -1 \\
1 & -1 & 1 & -1 & 0 & 1 \\
-1 & -1 & 1 & 1 & -1 & 0 \\
\end{array}
\right)\;.
\end{equation}
This is the same lattice as was found in \cite{Afkhami-Jeddi:2020ezh}, the generator matrices are identical. So we have found that the conjectural optimal Narain CFT with $c=7$ and $\Delta_1=\sqrt{4/3}$.
can be obtained from an
N-type code using construction \ref{constructionN}. We have checked, there are no other $[7,7,4]$ N-type codes besides this one.

\subsection{Other central charges}

Having found that the optimal Narain CFTs for $c=6,7$ are code CFTs, we move on to the case $c\neq 6,7$. As we now explain, we do not expect to be able to obtain an optimal Narain CFTs using construction \ref{constructionN} for any other values of $c$.

First we discuss $c<6$. The corresponding optimal Narain CFTs were oullined in \cite{Afkhami-Jeddi:2020ezh} (see table \ref{listofoptimal}). These CFTs cannot be reproduced from a code using construction \ref{constructionN}, since the spectral gap for any lattice obtained via construction \ref{constructionN} must take the form $\Delta_1=\frac{n}{2\sqrt 3}$ for some $n\in\mathbb{N}$, and this is not the case for any of the optimal CFTs for $c<8$, $c\neq 6,7$. For $c>7$, we again do not expect to be able to obtain optimal CFTs due to the bound $\Delta_1\leq 2/\sqrt 3$. This bound is saturated for $c=6,7$, while for $c=8$ the optimal CFT has $\Delta_1>2/\sqrt 3$. Since we expect $\Delta_1$ to be an increasing function of $c$, we conclude it would be impossible to construct any optimal $c>7$ CFT using the construction developed in this paper. 

 However, it is possible that other constructions can reproduce optimal CFTs at other values of $c$; indeed, the optimal CFTs with $c=3,4,5$ are related to codes and follow from the construction of \cite{Dymarsky:2020qom}. That construction  maps codes over $F_4$ to Narain CFTs, and for $c=3,4,5$ the optimal Narain lattices have the generator matrix  \eqref{eq:Lambda}, with $\frac{1}{\sqrt 2}\gamma^*=\sqrt 2\gamma=I$ and with the $B$ matrix corresponding to a fully connected graph on $c$ nodes, $B_{ij}=1$ for $i>j$. In general, we expect that all CFTs from the table \ref{listofoptimal} can be reproduced using codes; for example, the $c=8$ lattice is the Barnes-Wall lattice, which can be related to a code over $F_9$ \cite{SPLAG}. 

An important step would be to  develop a sequence of constructions which would yield optimal theories for larger values of $c$. On this path, the first step may not even be the explicit relation to codes, but a new algebraic ansatz for the partition function along the lines of \eqref{ZCFT}, amended with the analogs of the algebraic conditions (\ref{eq:T_transf},\ref{eq:mac}), which would ensure modular invariance. 

\section{Conclusions}\label{sec:conclusions}
In this paper we constructed a mapping from the family of codes over $F_4$ which satisfy the conditions of \ref{claim2}
and \ref{claim1} to the space of Narain theories. We call codes satisfying these conditions
N-type and the corresponding CFTs code theories. 
Starting from an N-type code we explicitly construct the Narain lattice and evaluate its theta-series, i.e.~the corresponding CFT torus partition function, in terms of the code's extended enumerator polynomial, see \eqref{eq:extended_enumerator} and \eqref{ZCFT}. Modular invariance of the partition function reduces to the algebraic identities \eqref{eq:mac} and \eqref{eq:T_transf}, which can be solved in terms of invariant polynomials, as discussed in section \ref{sec:invpol}.
Quite interestingly, the associated ring of invariant polynomials is not freely generated, in contrast to other known examples
\cite{nebe2006self}.
All N-type codes can be parametrized by a binary classical code, specified by matrix $b$, a pairwise permutation matrix $S$ and an antisymmetric matrix $\tilde B$ obeying \eqref{eq:gen_matrix_constraint}. 
As a result we could construct all codes of small length $n$ and found many interesting examples. In particular, we found that the conjecturally optimal Narain theory with $c=6$, based on the rescaled Coxeter-Todd  lattice, also known as $K_{12}$, understood as a Narain lattice \cite{Afkhami-Jeddi:2020ezh} is associated with the Hexacode, the unique $[6,6,4]$ self-dual code over $F_4$. Furthermore, the conjecturally optimal Narain theory with $c=7$ is associated with the ``septacode'' introduced in section \ref{sec:septacode}, the unique $[7,7,4]$ N-type code. 

Our construction is similar in spirit to previous works relating classical codes and chiral theories \cite{dolan1996conformal,dong1998framed,gaiotto2018holomorphic} as well as those relating quantum codes to Narain CFTs \cite{Dymarsky:2020qom,dymarsky2021solutions}. At the same time there are important novelties which we would like to emphasize. This is the first construction, to our knowledge, that maps codes to non-chiral {\it non-rational} theories. A natural question to ask is how general the relation between codes and CFTs could be. In all known examples of code theories the partition function is a sum of a handful of ``characters'' suggesting the theories in question are rational. Now we clearly see this is not necessarily the case. It thus remains an open question to chart the space of theories with the code counterparts, see \cite{upcoming}, and investigate if this relation can be extended beyond Narain theories. 

Among the code theories constructed in this paper are the conjecturally optimal Narain CFTs, i.e.~those with the largest value of spectral gap, for $c=6,7$. Together with the construction of \cite{Dymarsky:2020qom} we find that the optimal theories for $3\le c\leq 7$ are related to codes. This is likely to be true for $c=1,2$ and $c=8$ as well: in the former cases the partition function is a sum of ``characters'' and in the latter case the lattice is the Barnes-Wall lattice, which can be constructed using a code over $F_9$. More generally, any rational Narain CFT can be related to codes \cite{upcoming}, and this likely extends to any finite CFT \cite{Etingof:2002dn}. 
This observation prompts  the question if the optimal theories for larger $c$ could also be related to codes and if perhaps there are appropriate series of constructions which could describe optimal theories with an arbitrarily large 
central charge. This question can be reformulated in terms of the modular bootstrap program. The present construction, as well as that one of \cite{dymarsky2021solutions, Dymarsky:2020qom}, reduce the modular bootstrap constraints to algebraic constraints at the space of polynomials. While the present construction is somewhat nontrivial and involves polynomials of $14$ variables, the core idea remains the same. The modular bootstrap  becomes the question of identifying invariant polynomials and applying straightforward linear algebra. The question we would like to pose is to formulate a continuous or large discrete family of the appropriate Ans$\ddot{\rm a}$tze which would reduce the question of maximizing the spectral gap to the question of identifying the optimal algebraic identity and the ``characters'' such that the optimal partition functions for various $c$ would be given by some appropriate generalization of \eqref{ZCFT}. We hope to address the question of formulating such a meta-bootstrap approach in the future. 

\section*{Acknowledgements}
We thank N.~Angelinos, R.~Kalloor and M.~Watanabe for collaborations at the early stages of this project and M.~Buican and R.~Radhakrishnan  for discussions and reading the manuscript. A.D.~is grateful to Weizmann Institute of Science for hospitality and acknowledges  sabbatical support of the Schwartz/Reisman Institute for Theoretical Physics, and support by the NSF under grant PHY-2013812. A.S. is supported in part by an Israel Science Foundation center for excellence grant (grant number 2289/18), by the Minerva foundation with funding from the Federal German Ministry for Education and Research, and by grant no. 2018068 from the United States-Israel Binational Science Foundation (BSF).

\newpage

\begin{appendices}
	
\addtocontents{toc}{\protect\setcounter{tocdepth}{1}}

\section{Proofs}

\subsection{Isoduality of construction A lattice of codes over $F_4$}\label{app:isoduality}

We prove that
\begin{equation}
\tilde \Lambda^*(\C)=O_{\pi/2}^{(n)}\tilde \Lambda(\C^*)
\end{equation}
for a construction A lattice $\Lambda(\C)$ where $\C$ is a code over $F_4$ of length $n$.

For convenience, we define $\lambda(x)=(\text{Re}(x),\text{Im}(x))/3^{1/4}\in \mathbb{R}^{2n}$ for any $x\in F_4^n$. Then for example a vector $\alpha\in \Lambda(\C)$ takes the form $\alpha=\lambda(c+2n\cdot x)$ for $x\in F_4^n,n\in\mathbb{Z}^n$ and $c\in \C$. An important identity is that for $x\in F_4^n$,
\begin{equation}\label{eq:lambda_identity}
\lambda(x)O_{\pi/2}^{(n)}\lambda(y)\equiv \frac{1}{2}(x,y)\,\mod\, 1\;.
\end{equation}

First we prove 
\begin{equation}\label{eq:one_dir}
O_{\pi/2}^{(n)}\Lambda(\C^*)\subset \Lambda^*(\C)\;.
\end{equation}
Take $\alpha\in \Lambda(\C^*)$ of the form $\alpha=\lambda(c^*+2n\cdot  x)$ for $c^*\in \C^*$. Now take $\beta\in \Lambda(\C)$ of the form $\beta=\lambda(c+2m\cdot  y)$ for $c\in\C$. We must show that $(O_{\pi/2}^{(n)}\alpha)\cdot \beta \in\mathbb{Z}$. Using
\eqref{eq:lambda_identity}, we find
\begin{equation}
(O_{\pi/2}^{(n)}\alpha)\cdot \beta \equiv \frac{1}{2}(c^*,c)\,\mod 1\;.
\end{equation}
Since $c^*\in\C^*$, we must have $(c,c^*)\equiv 0 \,\mod 2$, and so $(O_{\pi/2}^{(n)}\alpha)\cdot \beta \in\mathbb{Z}$ as required.

Next we show 
\begin{equation}\label{eq:two_dir}
\Lambda^*(\C)\subset O_{\pi/2}^{(n)}\Lambda(\C^*) \;.
\end{equation}
Take $\alpha\in \Lambda^*(\C)$. Then since $\lambda(2x)\in \Lambda(\C)$ for any $x\in F_4^n$, we learn that $\alpha\cdot \lambda(2x)\equiv0\,\mod 1$. Considering $x$ such that all elements are zero apart from one, we learn that $\alpha$ must take the form $\lambda(c'+2n\cdot y)$ for $c',y\in F_4^n$. Now take $c\in \C$, then $\lambda(c)\in\Lambda(\C)$ and so 
\begin{equation}
 \lambda(c)O_{\pi/2}^{(n)} \alpha \equiv 0\mod 1\;,
\end{equation}
but on the other hand due to \eqref{eq:lambda_identity} we find
\begin{equation}
\frac{1}{2}(c,c')\equiv\lambda(c)O_{\pi/2}^{(n)} \alpha\mod 1\;,
\end{equation}
and so $(c,c')\equiv 0\mod\, 2\;,$ so $c'\in \C^*$. As a result, $\alpha\in O_{\pi/2}^{(n)}\Lambda(C^*)$. Combining \eqref{eq:one_dir} and \eqref{eq:two_dir} we learn that \begin{equation}
\Lambda^*(\C)= O_{\pi/2}^{(n)}\Lambda(\C^*)
\end{equation}
as required.

\subsection{Proof that a self-dual code with $w_{\omega}(c)-w_{\omegab}(c)=0\mod\,4$ is invariant under conjugation}\label{app:proof}

We want to prove the following: assume $\C$ is a self-dual code over $F_4$ and $\forall c\in \C$,
\begin{equation}
m(c)=w_{\omega}(c)-w_{\omegab}(c)=0\mod4\;.
\end{equation}
Here $w_{x}(c)$ is the number of letters in $c$
which are equal to $x\in F_4$. We want to prove that $\P(\C)=\C$. 

It is enough to show that if $c\in \C$ then for all $c_0\in \C$, we have
$(\overline{c},c_0)=0$ where $\overline c$ is the conjugate of $c$.
We denote by $w_x^y$ the
number of positions where $c$ has the letter $x$ and $c_{0}$ has
the letter $y$. Then for example
\begin{equation}
\sum_{y\in F_4}w_\omega^y=w_{\omega}(c)\;.\label{eq:1}
\end{equation}
So for example we have
\begin{equation}
\left(c,c_0\right)=w_\omega^1 +w_{\omegab}^1 +w_1^{\omega}+w_1^{\omegab}+w_\omega^{\omegab}+w_{\omegab}^{\omega}\equiv0\mod2
\end{equation}
and we want to show that
\begin{equation}\label{eq:to_show}
\left(\overline{c},c_0\right)= w_\omega^1 +w_{\omegab}^1 +w_1^{\omega}+w_1^{\omegab}+w_{\omegab}^{\omegab}+w_\omega^{\omega}\equiv0\mod2\;.
\end{equation}

The important point is that since $c+c_{0}\in \C$, we must have 
\begin{equation}
m\left(c_0+c\right)=w_{\omega}\left(c+c_0\right)-w_{\omegab}\left(c+c_0\right)\equiv0\mod4\;.
\end{equation}
Explicitly this means that
\begin{equation}
\left(w_0^{\omega}+w_\omega^{0}+w_{\omegab}^1 +w_1^{\omegab}\right)-\left(w_\omega^1 +w_1^{\omega}+w_{\omegab}^0+w_0^{\omegab}\right)\equiv0\mod4\;.
\end{equation}
Using the contraints of the form (\ref{eq:1}), we can write
\begin{align*}
w_\omega^0 & =w_{\omega}(c)-w_\omega^1 -w_\omega^{\omega}-w_\omega^{\omegab}\\
w_{\omegab}^0 & =w_{\omegab}(c)-w_{\omegab}^1 -w_{\omegab}^{\omega}-w_{\omegab}^{\omegab}\\
w_0^{\omega} & =w_{\omega}(c_0)-w_1^{\omega}-w_\omega^{\omega}-w_{\omegab}^{\omega}\\
w_0^{\omegab} & =w_{\omegab}(c_0)-w_1^{\omegab}-w_\omega^{\omegab}-w_{\omegab}^{\omegab}
\end{align*}
plugging this in we find
\begin{align*}
m(c_0+c)=& \left(w_{\omega}(c_0)-w_1^{\omega}-w_\omega^{\omega}-w_{\omegab}^{\omega}+w_{\omega}(c)-w_\omega^1 -w_\omega^{\omega}-w_\omega^{\omegab}+w_{\omegab}^1 +w_1^{\omegab}\right)\\
& -\left(w_\omega^1 +w_1^{\omega}+w_{\omegab}(c)-w_{\omegab}^1 -w_{\omegab}^{\omega}-w_{\omegab}^{\omegab}+w_{\omegab}(c_0)-w_1^{\omegab}-w_\omega^{\omegab}-w_{\omegab}^{\omegab}\right)\equiv0\mod4\;.
\end{align*}
Using $m(c)=w_{\omega}(c)-w_{\omegab}(c)=0\mod4$
and similarly for $m(c_0)$, this can be simplified to
\begin{equation}
m(c_0+c)=-2w_1^{\omega}-2w_\omega^{\omega}-2w_\omega^1 +2w_{\omegab}^1 +2w_1^{\omegab}+2w_{\omegab}^{\omegab}\equiv0\mod4\;,
\end{equation}
which we rewrite as
\begin{equation}
m(c_0+c)/2=w_1^{\omega}+w_\omega^1 +w_{\omegab}^1 +w_1^{\omegab}+w_\omega^{\omega}+w_{\omegab}^{\omegab}\equiv0\mod2\;.
\end{equation}
but comparing to \eqref{eq:to_show} we find that $m(c_0+c)/2\equiv (\overline{c},c_0)\mod 2$, and so
$(\overline{c},c_0)\equiv0\mod2$ as required.

\subsection{Proof of the expression for the dual binary subcode}\label{app:binary_1}

Consider an N-type code $\C$ with a binary subcode $\C_B$. Define
\begin{equation}
T=\{ S(c+\overline c)\, |\, c\in \C \}=\{ c^*+\overline {c^*}\, |\, c^*\in \C^* \}\;,
\end{equation}
where we used the fact that $\C^*=S(\bar \C)$. Then we would like to show that $T=\C_B^*$, where $\C_B^*$ is the dual of $\C_B$ with respect to the conventional binary inner product $(\,,\,)_B$. To show this, we will prove $T^*=\C_B$.

First show $T^*\subset \C$. Take $b\in T^*\in F_2^n,c^*\in \C^*$, then it is enough to show that $(b,c^*)=0$. Calculate:
	\begin{equation}
	\left(b,c^*\right)=(b, c^*)_B+(b,\overline{c^*})_B=(b,c^*+\overline{c^*})=0\;
	\end{equation}
	where we used $\overline b = b$ and the fact that $c^*+\overline{c^*}\in T$. So $T^*\subset \C$.	
	Next we show that all binary codewords in $\C$ are also in $T^*$. take binary $c\in \C$, and take $c^*+\overline{c^*}\in T$ for $c^*\in \C^*$. Then
	\begin{equation}
	(c,c^*+\overline{c^*})_B=(\overline{c},c^*)_B+(c,\overline{c^*})_B=\left(c,c^*\right)=0
	\end{equation}
	since $c^*\in \C^*$. So any binary $c\in\C$ is also in $T^*$. So we have proven that $T^*=\C_B$.

\subsection{Proof that $\C_B^*$ is even}\label{app:binary_2}

We now show that  $\C_B^*$ is even with respect to the inner product $(\cdot,\cdot)_{B,S}$ for an N-type code $\C$. Take $b\in \C_B$, then $b=S(c+\overline c)$ for $c\in \C$. Then
	\begin{equation}
	(b,b)_{B,S}=(c+\overline c,c+\overline c)_{B,S}=(c,S(c))+(c,S(\overline c))\;.
	\end{equation} 
	since $\C^\perp=S(\overline C)$ due to \ref{claim1}, the term $(c,S(\overline c))$ vanishes. We are left with showing that $(c,S(c))$ is 0 mod 2. Note that (using the notation around \eqref{eq:eveness_condition}),
	\begin{eqnarray}
		(c,S(c))\equiv \sum_i b_i b_{S(i)}-a_i a_{S(i)}\mod 2
	\end{eqnarray}
	which vanishes due to \eqref{eq:eveness_condition}, as required.

\section{The generator matrix}\label{app:generating_matrix}

The generator matrix for the Narain lattice is given by
\begin{equation}\label{eq:generating_matrix_general_construction}
\Lambda'=\twomatrix{\gamma^{*}}{\sqrt{3}B\gamma S}{}{\sqrt{3}S\gamma S}/3^{1/4}\;.
\end{equation}
This $\Lambda'$ is self-dual with respect to $g'$. Note that we can also write this as
\begin{equation}\label{eq:generating_matrix_general_construction_no_S}
\Lambda'=\twomatrix{\mathbb{1}_{n}}{}{}{S}\twomatrix{\gamma^{*}}{\sqrt{3}B\gamma }{}{\sqrt{3}\gamma }\twomatrix{\mathbb{1}_{n}}{}{}{S}/3^{1/4}\;.
\end{equation}

Now we can try to find the minimal form for $\gamma^*,B$. Note that in \eqref{eq:generating_matrix_general_construction_no_S}, we can ignore the matrix on the RHS since it is in $GL(2n,\mathbb{Z})$. We will now simplify the matrix by adding certain columns of the matrix to other columns, which amounts to multiplying on the right by elements of $GL(2n,\mathbb{Z})$. 

First, since the columns involving $\gamma^*$ correspond to elements of $\mathbb{Z}[\omega]$ which are only elements of $\mathbb{Z}$, we find that the columns of $\gamma^*$ $\mod\,2$ correspond to binary codewords. In addition, no other column can correspond to a binary codeword, and so all binary codewords can be generated by the columns of $\gamma^*$. We thus learn that $\gamma^*$ must be the generator matrix of the construction A lattice of the binary subcode $\C_B\subset \C$. 
As a result, $\gamma^*$ can be brought to the form
\begin{equation}
\gamma^*=\Lambda(\C_B^\perp)=\twomatrix{2\mathbb{1}_{n-k}}{b^T}{}{\mathbb{1}_{k}}\;,
\end{equation}
where $b^T$ is an $(n-k)\times k$ matrix. We immediately find (by definition) \begin{equation}
\gamma =(\gamma^{*-1})^T= \frac{1}{2}\twomatrix{\mathbb{1}_{n-k}}{}{-b}{2\mathbb{1}_{k}}\;.
\end{equation}
Next, since $B$ is antisymmetric we write it as
\begin{equation}
B=\left(\begin{array}{cc}
B_{11} & B_{12}\\
-B_{12}^{T} & B_{22}
\end{array}\right) \;.
\end{equation}
Finally, $S$ can be written as 
\begin{equation}
S=S^{T}=S^{-1}=\left(\begin{array}{cc}
S_{11} & S_{12}\\
S_{12}^{T} & S_{22}
\end{array}\right)\;.
\end{equation}
So the explicit form for the generator matrix $\tilde \Lambda$ is
\begin{equation}
\Lambda^{\prime}=\left(\begin{array}{cc}
\gamma^{*} & B\gamma\\
& \sqrt{3}S\gamma
\end{array}\right)=\left(\begin{array}{cccc}
2\mathbb{1}_{n-k} & b^{T} & \frac{1}{2}\left(B_{11}-B_{12}b\right) & B_{12}\\
& \mathbb{1}_{k} & -\frac{1}{2}\left(B_{12}^{T}+B_{22}b\right) & B_{22}\\
&  & \frac{\sqrt{3}}{2}\left(S_{11}-S_{12}b\right) & \sqrt{3}S_{12}\\
&  & \frac{\sqrt{3}}{2}\left(S_{12}^{T}-S_{22}b\right) & \sqrt{3}S_{22}
\end{array}\right)\;.
\end{equation}
Consider some column on the far right hand side in terms of elements of $\mathbb{Z}[\omega]$. Since the elements of $S$ are integers, each element of the column must take the form $c+2\omega$ for some $c\in\C_B$. But then we can add the columns of $\gamma^*$ to this vector to being every element to the form $2\omega$. This amounts to setting $B_{12}=S_{12}$ and $B_{22}=S_{22}$. Next, since $B_{22}$ is now an integer matrix, we use the colums of $\gamma^{*}$ to set $B_{22}$ to zero by adding 
$\left(\begin{array}{c}
-b^{T}S_{22}\\
-S_{22}
\end{array}\right)$ to $\left(\begin{array}{c}
B_{12}\\
B_{22}
\end{array}\right)$,\footnote{It is crucial that $S_{22}$ is an integer matrix, since as a result this amounts to multiplying on the right by an element of $GL(2n,\mathbb{Z})$.} so we get
\begin{equation}
\Lambda^{\prime}=\left(\begin{array}{cc}
\gamma^{*} & B\gamma\\
& \sqrt{3}S\gamma
\end{array}\right)=\left(\begin{array}{cccc}
2\mathbb{1}_{n-k} & b^{T} & \frac{1}{2}\left(B_{11}-B_{12}b\right) & S_{12}-b^{T}S_{22}\\
& \mathbb{1}_{k} & -\frac{1}{2}\left(B_{12}^{T}+B_{22}b\right) & 0\\
&  & \frac{\sqrt{3}}{2}\left(S_{11}-S_{12}b\right) & \sqrt{3}S_{12}\\
&  & \frac{\sqrt{3}}{2}\left(S_{12}^{T}-S_{22}b\right) & \sqrt{3}S_{22}
\end{array}\right)\;.
\end{equation}
Before moving on we will use the columns $\left(\begin{array}{c}
2\mathbb{1}_{n-k}\\
0
\end{array}\right)$ to flip the sign of the final colums, so we have
\begin{equation}
\Lambda^{\prime}=\left(\begin{array}{cc}
\gamma^{*} & B\gamma\\
& \sqrt{3}S\gamma
\end{array}\right)=\left(\begin{array}{cccc}
2\mathbb{1}_{n-k} & b^{T} & \frac{1}{2}\left(B_{11}-B_{12}b\right) & b^{T}S_{22}-S_{12}\\
& \mathbb{1}_{k} & -\frac{1}{2}\left(B_{12}^{T}+B_{22}b\right) & 0\\
&  & \frac{\sqrt{3}}{2}\left(S_{11}-S_{12}b\right) & \sqrt{3}S_{12}\\
&  & \frac{\sqrt{3}}{2}\left(S_{12}^{T}-S_{22}b\right) & \sqrt{3}S_{22}
\end{array}\right)\;.
\end{equation}
Next, consider the columns of 
\begin{equation}
\left(\begin{array}{c}
\frac{1}{2}\left(B_{11}-B_{12}b\right)\\
-\frac{1}{2}\left(B_{12}^{T}+B_{22}b\right)\\
\frac{\sqrt{3}}{2}\left(S_{11}-S_{12}b\right)\\
\frac{\sqrt{3}}{2}\left(S_{12}^{T}-S_{22}b\right)
\end{array}\right)\;.
\end{equation}
Due to the general constraints on the column vectors \eqref{eq:evvecs}, this row obeys
\begin{align}
\frac{1}{2}\left(B_{11}-B_{12}b\right)	&=\frac{1}{2}\left(S_{11}-S_{12}b\right)+Y_{1}\;,\\
-\frac{1}{2}\left(B_{12}^{T}+B_{22}b\right)	&=\frac{1}{2}\left(S_{12}^{T}-S_{22}b\right)+Y_{2}\;. 
\end{align}
We once again use the columns of $\gamma^*$ to simplify this by adding $\left(\begin{array}{c}
-b^{T}Y_{2}\\
-Y_{2}
\end{array}\right)$ to $\left(\begin{array}{c}
\frac{1}{2}\left(B_{11}-B_{12}b\right)\\
-\frac{1}{2}\left(B_{12}^{T}+B_{22}b\right)
\end{array}\right)$,\footnote{again, it is crucial that $Y_{2}$ is an integer matrix so that this amounts to multiplying on the right by an element of $GL(2n,\mathbb{Z})$.} and we get
\begin{equation}
\Lambda^{\prime}=\left(\begin{array}{cccc}
2\mathbb{1}_{n-k} & b^{T} & \frac{1}{2}\left(B_{11}-B_{12}b\right)-b^{T}Y_{2} & b^{T}S_{22}-S_{12}\\
& \mathbb{1}_{k} & \frac{1}{2}\left(S_{12}^{T}-S_{22}b\right) & 0\\
&  & \frac{\sqrt{3}}{2}\left(S_{11}-S_{12}b\right) & \sqrt{3}S_{12}\\
&  & \frac{\sqrt{3}}{2}\left(S_{12}^{T}-S_{22}b\right) & \sqrt{3}S_{22}
\end{array}\right)\;.
\end{equation}
We have thus effectively set $B_{22}=0$, and we are close to setting $B_{12}=b^{T}S_{22}-S_{12}$. To finish, write $\frac{1}{2}\left(B_{11}-B_{12}b\right)-b^{T}Y_{2}$ explicitly:
\begin{align}
\frac{1}{2}\left(B_{11}-B_{12}b\right)-b^{T}Y_{2}	&=\frac{1}{2}\left(B_{11}-B_{12}b\right)-b^{T}\left(-\frac{1}{2}\left(B_{12}^{T}+B_{22}b\right)-\left(\frac{1}{2}\left(S_{12}^{T}-S_{22}b\right)\right)\right)\\
&=\frac{1}{2}\left[B_{11}-B_{12}b+b^{T}B_{12}^{T}+b^{T}B_{22}b+b^{T}S_{12}^{T}-b^{T}S_{22}b\right] \;.
\end{align}
Define the antisymmetric matrix
\begin{equation}
\tilde{B}=B_{11}-B_{12}b+b^{T}B_{12}^{T}+b^{T}B_{22}b\;,
\end{equation}
then we find
\begin{equation}
\frac{1}{2}\left(B_{11}-B_{12}b\right)-b^{T}Y_{2}= \frac{1}{2}\left(\tilde B+b^TS_{12}^T-S_{12}b+S_{12}b-b^{T}S_{22}b\right)\;.
\end{equation}
This finally allows us to set $B_{12}=b^{T}S_{22}-S_{12}$, in which case
\begin{equation}
\frac{1}{2}\left(B_{11}-B_{12}b\right)-b^{T}Y_{2}= \frac{1}{2}\left(\tilde B+b^TS_{12}^T-S_{12}b-B_{12}b\right)
\end{equation}

To summarize, our final result is
\begin{equation}
\Lambda^{\prime}=\left(\begin{array}{cc}
\gamma^{*} & B\gamma\\
& \sqrt{3}S\gamma
\end{array}\right)=\left(\begin{array}{cccc}
2\mathbb{1}_{n-k} & b^{T} & \frac12(\tilde B+b^TS_{12}^T-S_{12}b-B_{12}b) & B_{12}\\
& \mathbb{1}_{k} & -\frac{1}{2}B_{12}^T & 0\\
&  & \frac{\sqrt{3}}{2}\left(S_{11}-S_{12}b\right) & \sqrt{3}S_{12}\\
&  & \frac{\sqrt{3}}{2}\left(S_{12}^{T}-S_{22}b\right) & \sqrt{3}S_{22}
\end{array}\right)
\end{equation}
where $B_{12}=b^{T}S_{22}-S_{12}\;$. Comparing to the general form we started with, we find that we can always bring $B$ to the form
\begin{equation}
B=\frac{1}{\sqrt{3}}\left(\begin{array}{cc}
\tilde B+b^TS_{12}^T-S_{12}b & b^{T}S_{22}-S_{12}\\
S_{12}^{T}-S_{22}b & 0
\end{array}\right)\;.
\end{equation}
with $\tilde B$ an integral antisymmetric matrix.

\end{appendices}

\printbibliography

\end{document}